\documentclass[pre,aps,amsmath,amssymb,floatfix,preprint]{revtex4-1}
\usepackage[export]{adjustbox}
\usepackage{amssymb}
\usepackage{graphicx}% Include figure files
\usepackage{dcolumn}% Align table columns on decimal point
\usepackage{bm}
\usepackage{subfigure}
\usepackage[utf8]{inputenc}
\usepackage{color}
\usepackage{hyperref}
\hypersetup{colorlinks=true, citecolor=blue,
  urlcolor=black,
  pdfcreator={pdflatex},
}

\begin{document}
\title{On the build-up of effective hyperuniformity from large globular colloidal aggregates}

\author{Antonio Díaz-Pozuelo$^1$ and  Diego González-Salgado$^{3}$, and Enrique Lomba$^{1,2}$}
\email{Corresponding author: e.lomba@iqf.csic.es}

\affiliation{
$^1$Instituto de Qu\'{i}mica F\'{i}sica Blas Cabrera,
CSIC, Madrid, Spain\\
$^2$Grupo NAFOMAT, Facultade de Física, Universidade de Santiago de
Compostela, Spain\\
$^3$Modelización y Simulación de Materiales Nanoestructurados,
Universidad de Vigo, Unidad Asociada al CSIC por el IQF,  Departamento
de Física Aplicada, E-32004, Ourense, Spain}

\begin{abstract}
A simple three-dimensional model of a fluid whose constituent particles interact via a
short range attractive and long range repulsive potential is used to
model the aggregation  into large spherical-like clusters made up of hundreds of
particles. The model can be thought of as a straightforward rendition of colloid
flocculation into large spherical aggregates. We illustrate how temperature and
particle density influence the cluster size distribution and 
affect inter- and intra-cluster dynamics. The system is shown to exhibit
two well separated length and time scales, which can be tuned by the
balance between repulsive and attractive forces. Interestingly,
cluster aggregates at moderate/low temperatures approach a cluster glassy
phase whereas cluster particles retain a local liquid-like structure.
These states present  
a strong suppression of density fluctuations for a significant range of
relatively large wavelengths, meeting the criterion of effective
disordered hyperuniform materials as far as the intercluster structure is
concerned.

\end{abstract}

\maketitle
\section{Introduction}
Spontaneous pattern formation and self-limiting association can be
deemed  key phenomena both in 
biological and soft matter physics\cite{Stradner2004}. In the former, colloidal suspensions are
characteristic examples in which a variety of patterns have been
identified in systems as disparate as  amphiphilics\cite{Ciach2013},
and proteins such as lysozime\cite{Liu2010} or
antibodies\cite{Yearley2014} (see Kovalchuk and
coworkers\cite{Kovalchuk2009} and references therein for a more
comprehensive review of experimental examples).  As a particularly interesting illustration of
self-limiting aggregation, recently Sweatman and Lue
\cite{Sweatman2019} have argued that the formation of giant 
clusters due to the competition between attractive and repulsive
interactions might well be behind the condensation of proteins/nucleic
acids into  membraneless
organelles  within the nuclei of eukaryotic cells
\cite{Hirose2022}. This is an alternative view to the ``traditional''
liquid-liquid phase separation picture that is common among the
structural biology community\cite{Murthy2020}. 

As to pattern
formation in soft matter systems, there is a growing interest on this
subject, in particular for the possibilities that non-templated
thermodynamically controlled
nano-patterning\cite{Sear1999,Boles2016,Barad2021} offers in the field of
nanotechnology. In this connection, simple models
that exhibit short range attraction and long range repulsion (SALR),
\cite{Andelman1987,Seul1995} have shown to be able to reproduce the
experimental behavior in great detail. In the pioneering
work of Andelman and coworkers it was  predicted by means of a mean field
theory that, for a sufficiently large amplitude of repulsive
interactions, modulated cluster phases will emerge\cite{Archer2007}. The
central role of attraction/repulsion competition in the build-up of globular 
bubble cluster, and  ordered modulated phases or bicontinuous
percolating clusters that span up to macroscopic sizes, has been known for more than three
decades\cite{Seul1995}. The shape and size of the
clusters and their spatial organization is directly
determined by the balance between attractive
and repulsive forces, as shown by  the ground state calculations of Mossa et
al.\cite{Mossa2004}. Obviously, thermodynamic conditions (density and
temperature/pressure) play a key role as well.  Recently Liu and Xi \cite{Liu2019} proposed a
classification of SALR potentials in terms of the ratio between
repulsive and attractive forces and particle size. Thus SALR potential
are classified into type-I (range of the attraction
is less than 20\% of the particle size, and range of the repulsion is a fraction
of the particle size), type-II (range of the attraction less than 20\%
of the particle size, range of repulsion is comparable or larger than
particle size) and type-III (range of attraction beyond 20\% of
particle size and range of repulsion as in type II). In
Ref.~\cite{Liu2019} it was shown  that these three classes of interactions
lead to specific features as to their phase behavior and cluster
morphology. 

The phase behavior of SALR systems has been dealt with 
extensively using mean field and density functional theory approaches
\cite{Archer2007,Archer2008,Archer2008a,Bomont2012,Bomont2012a,Sweatman2014,Edelmann2016,Pini2017},
effective field theories\cite{Ciach2008b,Ciach2013,Konigslow2013}, and simulation
approaches \cite{Archer2007a,Klix2010,Godfrin2014,Zhuang2016,Zhuang2016a}. A 
comprehensive summary  
of relevant contributions in this connection can be found  in recent review works 
\cite{Liu2019,RuizFranco2021}. It should be stressed that in all these instances, the
long range repulsive component of the interaction is essential for
the self-limitation of the aggregation process, preempting
condensation/demixing transitions. The latter would drive the system
towards  spatially separate
macroscopic phases\cite{Godfrin2014} instead of the microstructuring
induced by clustering. It is 
worth mentioning that other
alternative approaches can explain self-limitation in aggregation processes as
well. Such is the case, 
for instance, when the presence of specific interactions in mixtures
limits  the growth of the aggregate due the saturation of
bonding sites by one of the components\cite{Palaia2022}, in parallel
with the chain termination in radical polymerization reactions. 

Finally, as stated by Klix and coworkers \cite{Klix2010} SALR systems
are intrinsically frustrated, and this also affects their dynamics,
which is mostly controlled either by mesoscopic order or by metastable
disorder\cite{Klix2010}. These authors actually identify three
dynamically arrested phases in their SALR model with long range
electrostatic repulsions: a Wigner glass
(moderate packing fraction/weak attractive interactions), a cluster
glassy state (low packing fraction, strong attractive interactions),
and a percolating gel phase (high packing fraction/strong attractive
interactions). The possibility of the existence of a Wigner glass phase
in disordered systems with long range electrostatic interactions was
first postulated by Bose and Wilke using mode coupling
theory\cite{Bosse1998}. Dawson et al \cite{Dawson2002} further
analyzed in detail the possibility of dynamically arrested states in
colloids stemming from either attractive or repulsive forces, whose ratio, as
 Klix et al. \cite{Klix2010} have shown, controls both the type of dynamics and
 topology of the glassy states.

 Experimental evidence of colloidal
 glassy states is  extensive and, for instance, numerous examples can
 be found in the review chapter by Weitz \cite{Weitz2011}. These
 glassy states are particularly interesting from the technological
 standpoint, since they provide an avenue to manufacture a wide range
 of solid-like materials starting from colloidal solutions, and as
 will be shown in this work, they  offer a feasible alternative  for the fabrication
 of disordered hyperuniform materials. These systems are exotic
 states, lying in between crystals and
 fluids\cite{Torquato2003,Torquato2018a}, i.e. while being
 structurally disordered, they exhibit a hidden order that, in common
 with crystalline phases, suppresses large scale density
 fluctuations. Aside from its ubiquitous occurrence in physical and
 biological systems (cf. see  Ref.~\cite{Torquato2018a}
 for a comprehensive review of the multiple systems in which
 disordered hyperuniformity is present), hyperuniform materials have been shown to
 display particularly interesting
 optical\cite{Florescu2009,Florescu2009a,Froufe-Perez2017,Zhou2019,Milosevic2019} and
 acoustic properties\cite{RomeroGarcia2019,Cheron2022}. Manufacturing
 hyperuniform materials from colloidal aggregates opens multiple paths
 to the fine tuning of their optical/acoustical properties, since in
 addition to the aggregate composition (and corresponding form factor of the
 constituent particles), the size and/or topology of the particulate
 aggregates can also be controlled by chemical and physical means.

In this article we will revisit a particularly simple model of
colloidal SALR system that can yield
all types of cluster phases, and ultimately will be shown to be
capable of producing disordered effective hyperuniform states. The
two dimensional version of this model was explored in detail
two decades ago by Imperio and Reatto 
\cite{Imperio2004,Imperio2006,Imperio2006a,Imperio2007}. The model, in
which both attractive and repulsive interactions are simple
exponentials, i.e. Kac interactions\cite{Kac1959}, was initially
introduced in this context by Sear and coworkers\cite{Sear1999} to account for the
spontaneous patterning of quantum dots at the air-water
interface. Later, Archer and coworkers \cite{Archer2007,Archer2007a,Archer2008a} used density
functional theory and extensive Monte Carlo simulations in a
three-dimensional version of the same model, in order to explore its
phase behavior and map its corresponding phase diagram. Schwanzer and
Kahl\cite{Schwanzer2010}  analyzed the competition between
clustering and vapor liquid condensation, and the cluster/particle
dynamics\cite{Schwanzer2016} again in two dimensions.  Interestingly,
the same Kac-potential model was also used by Meyra  
and coworkers \cite{Meyra2012,Meyra2015} to explain the formation of
vegetation patterns in environments with limited resources. This illustrates how
SALR effective interactions can provide a qualitative (and even
quantitative) explanation for spontaneous patterning over various
orders of magnitude in the spatial (and to some extent temporal)
dimension.   In
addition to all these studies in the bulk,
Bores et al.~\cite{Lomba2014a,Bores2015} performed simulations and
theoretical studies on the influence of confinement into disordered
media on the pattern formation for the same SALR model in two
dimensions as well.

In the comprehensive work of
Archer and Wilding \cite{Archer2007a}, in addition to the vapor-liquid
transition, two first order transitions were identified depending on the
packing fraction. Namely, a phase change at high dilution between a
vapor and a fluid of liquid-like spherical clusters (globular cluster phase), and another
transition between a liquid and fluid of spherical voids (bubble
cluster phase).  In this paper we will focus solely on the first
instance, with interaction parameters and thermodynamic
conditions tuned to yield relatively  large clusters.
    Here, using large scale
simulations we  have  analyzed the cluster size
distributions, the intercluster and intracluster microscopic
structure and dynamics illustrating their dependence on the
thermodynamic conditions. In contrast with the two-dimensional model, in
which as temperature is lowered both the intracluster and
intercluster structure rapidly freeze into  hexagonal arrangements, the
three-dimensional system retains the liquid-like structure of the
cluster droplets, and the freezing of intercluster structure is
frustrated, with the system displaying features characteristics of a
glass. In this work, we will exploit  the  long range of the net
inter-cluster interactions and relatively large attractive
interparticle interaction (a type-III SALR potential), to find the
conditions in which  clusters end-up in the vicinity 
of a dynamically arrested state, a
cluster glassy phase\cite{Klix2010}.  It will be
shown that these
states display a  strong
attenuation of density fluctuations for a significant region of small
wavenumbers. The system effectively resembles
a  stealthy disordered hyperuniform
material\cite{Torquato2003,Batten2008,Morse2023}, i.e. due to the
 suppression of density fluctuations for a certain region of large
 wavelengths, the 
material becomes to some extent ``invisible'' to the corresponding
radiation probes.

The rest of the paper is sketched as follows. In the next section we
will describe the SALR model and the simulation conditions. An
analysis of the onset of clustering is presented in Section \ref{onset}. The single
cluster structure (cluster size distributions, density profiles,
average life times) and dynamics  will be reviewed in Section
\ref{oneb}.  Two-particle and 
intercluster correlations, and the analysis of 
hyperuniformity are commented upon in Section
\ref{twob}. The article is closed with a summary of 
our most relevant conclusions and future prospects.

\begin{figure}[t]
%\begin{center}
\centering
\includegraphics[width=14cm,clip]{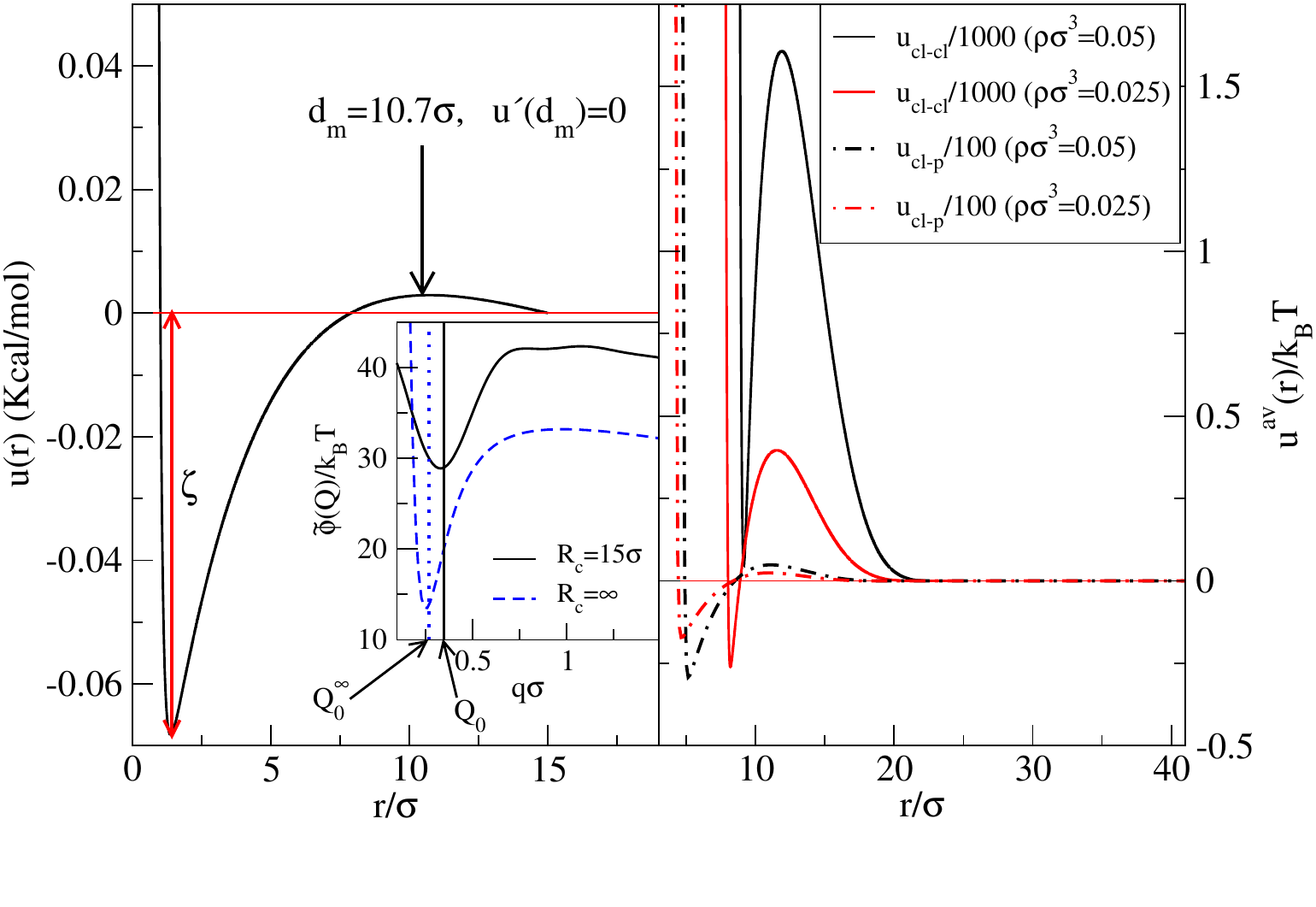}
   \caption{(Left) SALR interaction potential as described by
     Eq.~(\ref{pot}). In the inset the Fourier transform of non-singular part of the
     potential, $\tilde{\phi}(Q)$, (black  solid curve) is also represented. For
     comparison in dashed blue the corresponding transform of the
     untruncated potential is also plotted. In vertical lines we
     indicate the minima of both transforms which are the
     characteristic wave vectors for the truncated and untruncated
     interaction, namely $Q_0$ and $Q_0^\infty$. (Right) Average intercluster potential (solid 
     curves)  and cluster-particle potential
    (dashed curves), computed assuming an uniform average cluster
    density and spherical cluster radii.}
\label{ur}
%\end{center}
\centering
\end{figure}

 \begin{figure}[t]
\centering
\includegraphics[width=8cm,clip]{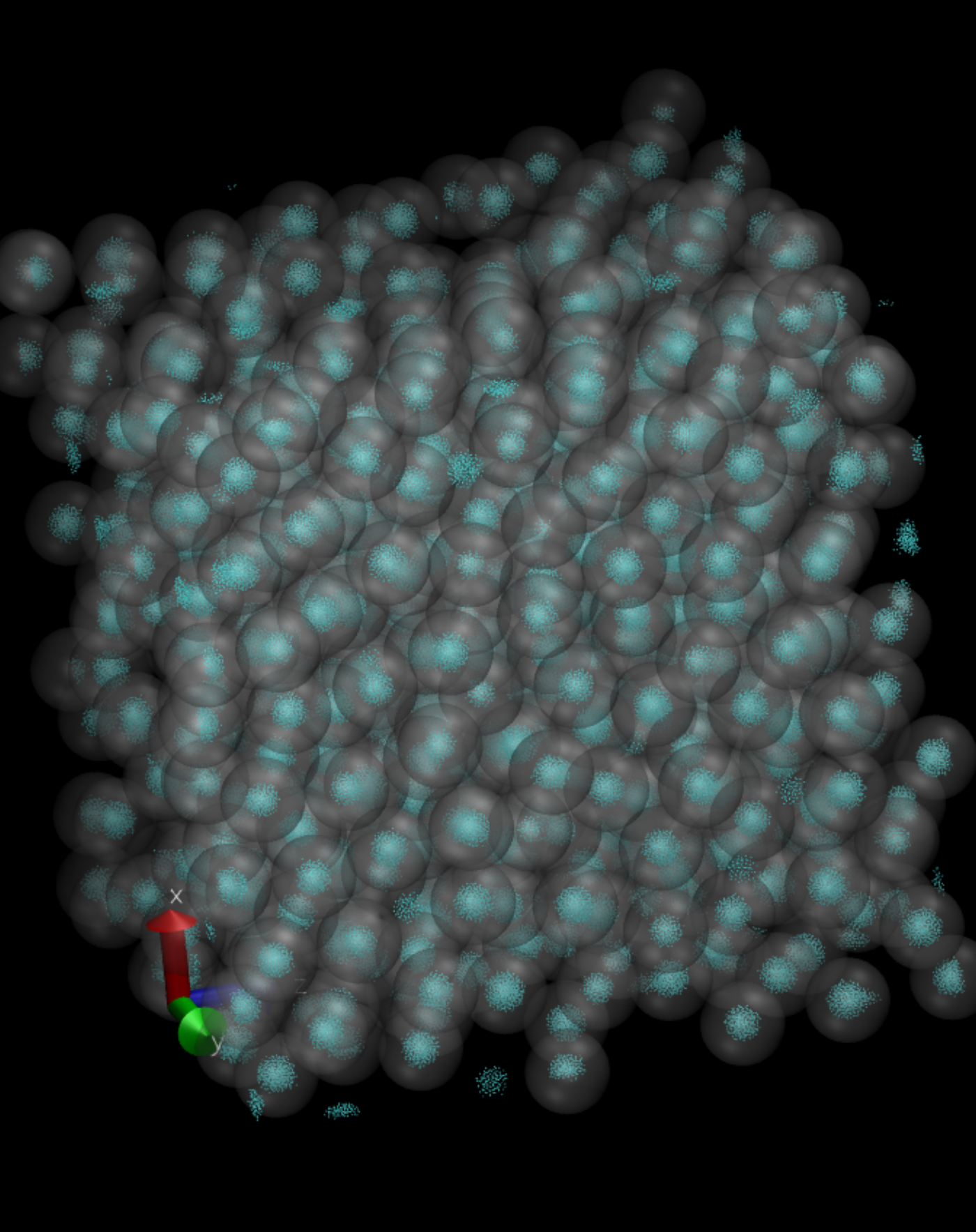}
   \caption{Snapshot of a cluster configuration in the globular phase
     for $\rho\sigma^3=0.05$, $k_BT/\zeta=3$. Transparent spheres
     represent the excluded volume around the clusters (approx. twice
     the particle cluster diameter) resulting from
     the long range interactions. The effective cluster density
     inferred from the excluded volume is
     $\rho_C\lambda_0^3=0.82$, corresponding to a rather dense fluid. \label{clus-conf}}
\end{figure}

\section{Model and methods\label{model}}
\begin{figure}[t]
\centering
\includegraphics[width=12cm,clip]{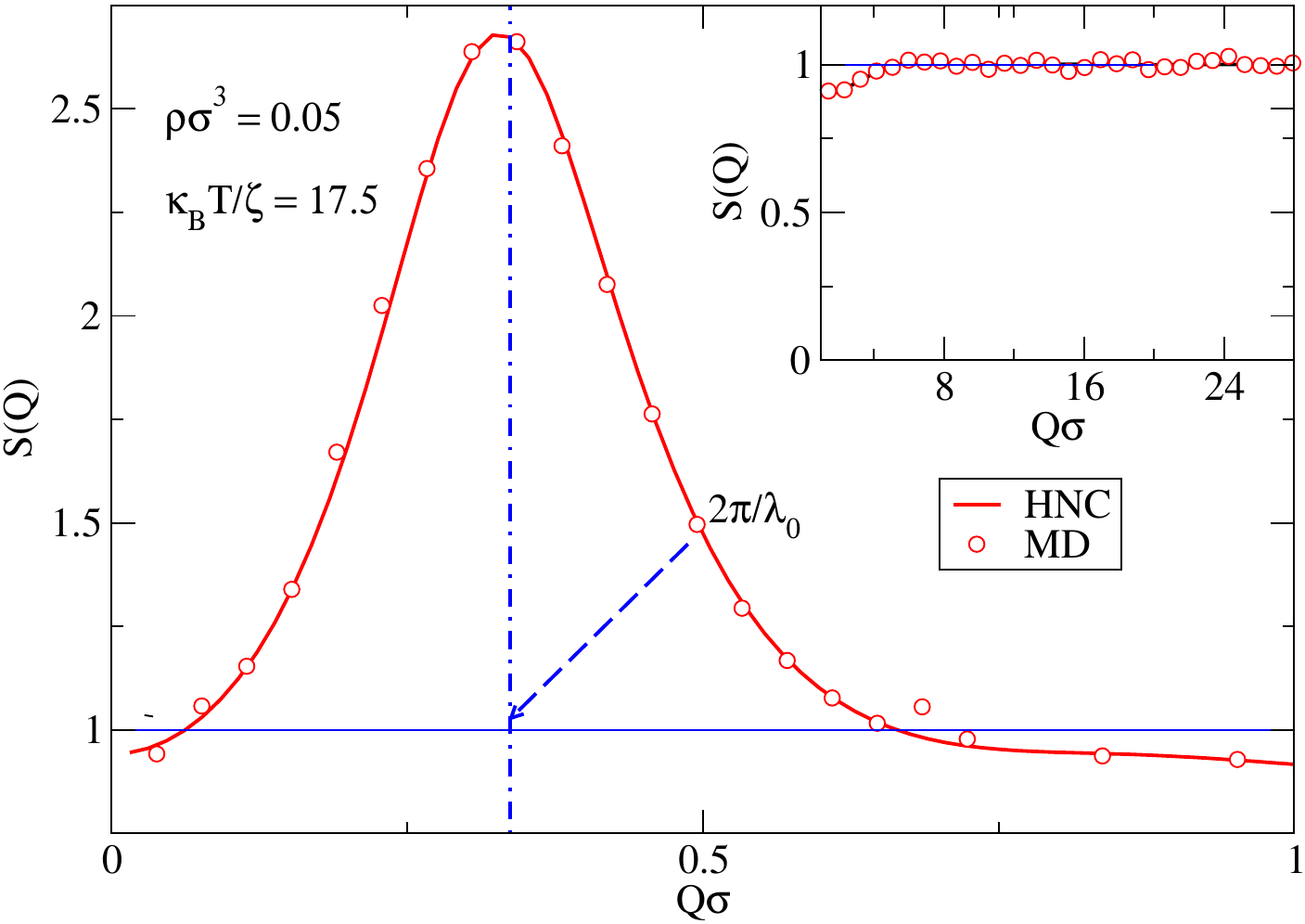}\label{sqHNC}
   \caption{High temperature structure factor from our HNC
     calculations (lines) vs MD results (symbols). In these conditions
     the system is in the homogeneous phase displaying the initial
     formation of aggregates. A vertical line
     marks the position of the characteristic wave vector, $2\pi/\lambda_0$, The inset depicts the medium range structure
     factor, wherein intracluster correlations should appear. The
     values close to one imply absence of correlations, i.e. ideal
     gas behavior.\label{sqhnc}}
\end{figure}

As mentioned in the Introduction the particles in our model will be
interacting via a two-exponential Kac potential of the form used in
Refs.~\cite{Sear1999,Imperio2004,Archer2007,
  Archer2007a,Bomont2012,Lomba2014a} to which we have 
added a soft core as in Ref.\cite{Lomba2014a} in order to
facilitate the use of molecular dynamics, namely  
\begin{equation}
u^{SALR}(r) = \varepsilon\left[K_re^{-\alpha_r r/\sigma}-K_ae^{-\alpha_a
    r/\sigma} +\left(\frac{\sigma}{r}\right)^{10}\right].
\label{SALR}
\end{equation}
Here $\sigma$ is a measure of the inner particle core diameter and
will be used as unit length. Since
this is a rather soft potential, the actual size, as estimated from
the particle-particle distribution function is $\sigma_{eff}\approx
0.8\sigma$.   As
mentioned by Sear and coworkers\cite{Sear1999}, while the use of the 
attractive exponential can be justified as a more or less crude
modelization of dispersive forces as in the classical works by Kac et
al.\cite{Kac1959,Kac1963}, the repulsive exponential is used solely
for the sake of
computational simplicity. Thus, with $K_a, K_r >0$, and $\alpha_a>\alpha_r>0$ one has the
characteristics of a SALR potential with the presence of a maximum for
a distance $d_m$, larger than that corresponding to the potential
minimum, and whose analytical expression reads
\begin{equation}
\label{r0}
d_m = \frac{\sigma}{\alpha_a-\alpha_r}\log
\frac{\alpha_aK_a}{\alpha_rK_r}.
\end{equation}
The effect of the inner repulsive core on the value of $d_m$ is
negligible.  $d_m$ is a very significant quantity  that in
the case of globular clusters approximately determines   the cluster size,
being a rough measure of its diameter. As will be
explained below, due to the long range character of the interaction,
the  average intercluster distance is approximately twice as large,
i.e. $\lambda_m=2d_m$. For
computational convenience  the
potential has been truncated and shifted at $R_c=60\sigma$, i.e., we
will be dealing with an interaction given by
\begin{equation}
\label{pot}
u(r) = \left\{\begin{array}{cc}
u^{SALR}(r) - u^{SALR}(R_c) & \mbox{if} \; r \le R_c \\
0 & \mbox{if} \; r > R_c 
\end{array} \right.
\end{equation}
Setting $K_r=1$ and
$K_a=2$, $\sigma=4\;$\AA, $\varepsilon = 0.1$ kcal/mol, $\alpha_r=0.1$ and
$\alpha_a=0.25$ one gets the potential depicted in the left graph of
Figure \ref{ur}. 

Alternatively, one can resort to  the characteristic wave
vector of the interaction (and its corresponding correlation length) which
is the absolute minimum of the Fourier transform of its non-singular part, namely, the potential
without the strong short range repulsion, $\phi(r) = u(r)-
\varepsilon\left(\frac{\sigma}{r}\right)^{10}$. Note that for
$r>2\sigma$ $u(r)$ and $\phi(r)$ are practically identical.   Its connection with
the structure factor and the build up of intermediate range order can
easily be understood recalling that 
\begin{equation}
\lim_{Q\rightarrow 0} S(Q) = 1+\rho\tilde{h}(Q) =
\left[1-\rho\tilde{c}(Q)\right]^{-1} \approx
  \left[\chi_0^{-1}+\rho\tilde{\phi}(Q)/(k_BT)\right]^{-1}
\label{msa}
\end{equation}
where the tilde denotes a Fourier transform, $\rho$ is the particle
number density,  $k_B$ is 
Boltzmann's constant, and we have made use of the Ornstein-Zernike
relation\cite{HansenBook3rd} that connects the pair distribution
function $g(r) = h(r)-1$ with the direct correlation function,
$c(r)$. Additionally, $\chi_0$ represents the isothermal compressibility of a
reference fluid of particles interacting with the 
repulsive short range part of the potential ($\propto r^{-10}$ in our case). This latter quantity stems
from  the random phase
approximation (cf Ref.~\cite{Archer2008a}) which gives 
$c(r) \approx c_0(r) -
u(r)/k_BT\approx c_0(r)-\phi(r)/k_BT$, being $c_0(r)$  the direct
correlation function of the reference fluid. For our densities of interest
$\chi_0\approx 1$. One
sees that a minimum in the potential's Fourier
transform present at $Q_0$ implies a maximum at $S(Q_0)$. If the
denominator in the r.h,s. of (\ref{msa}) vanishes, the system reaches a Lifshitz point,
corresponding to the onset of infinitely long ranged modulated phases. In our case one can
easily determine the position of the minimum for the untruncated interaction
(\ref{SALR}), as
\begin{eqnarray}
Q_0^\infty &=& \sqrt{\frac{c\alpha_a^2-\alpha_r^2}{1-c}},\;\;\;c =
\left(\frac{\alpha_rK_r}{\alpha_aK_a}\right)^{1/3}\nonumber\\
\lambda_0^\infty & = & 2\pi/Q_0^\infty,
\label{qm}
\end{eqnarray}
being $\lambda_0^\infty$ being the correlation length for the
  system with untruncated interaction. In the inset of the left graph
  of Figure \ref{ur} one sees clearly the 
presence of a marked minimum in the untruncated potential at
$Q_0^\infty\sigma = 0.253$.  Due to truncation this minimum is shifted
in our case to $Q_0\sigma = 0.337$. Remarkably,
the correlation value estimated from $d_m$ (i.e. the potential maximum), namely, $2\pi/\lambda_m =
0.293$ is relatively close. This is simply due to fact that the lowest
order approximation for the value of both
quantities (in particular for $Q_0^\infty\sigma$) is $\alpha_a$, (here 0.25). This only applies to the specific
functional form of our interaction and our choice of parameters
($\alpha_aK_a>>\alpha_rK_r$). 

Now, in lamellar phases, $\lambda_0$
will determine the modulation distance, and in the bubble phase the
average distance between the bubbles. In all cases the signature of
this correlation length, $\lambda_0$, is the presence of a high intensity peak in the
structure factor, S(Q), located at $Q_0$.  With our choice of
parameters $d_m=10.7\sigma$. The corresponding estimate from the
characteristic wavevector $Q_0$
gives a distance $d_0=9.32\sigma$ relatively close to the former,  The
cluster diameter will lie somewhere between 9 and 
11 molecular diameters. This will be further
confirmed in the next Section when discussing the average cluster
density profiles. We
will consider systems at relatively low 
number densities in order to guarantee the formation of micellar-like phases.
Two number densities will be studied, $\rho\sigma^3 = 0.025$, and 
0.05. As expected from the difference between $Q_0^\infty$ and $Q_0$,  the truncation of
the SALR potential implies that cluster sizes will be somewhat smaller
than those corresponding to an untruncated potential, also the
corresponding correlation lengths will also be
slightly smaller.

The fact that the $\lambda_0$ is approximately twice the position of the
potential maximum is also easily understood in terms of average
  cluster-cluster interactions in the case of globular
clusters.  In this instance one can assume a uniform effective density, $\rho_{cl}^{eff}$,
inside the spherical clusters of effective radius $R_{cl}^{eff}$.
These two quantities can be considered as optimization parameters in a
Reverse Monte Carlo procedure\cite{McGreevy2001}, in which the
interaction potential is computed from 
\begin{equation}
u_{cl-cl}^{av}(r;\rho_{cl}^{eff},R_{cl}^{ef}) = \left(\rho_{cl}^{eff}\right)^2\int_{V_{cl}} d{\bf
  r}_1\int_{V_{cl}} d{\bf r}_2 u(|{\bf r}-{\bf r}_1+{\bf r}_2|)
\label{uav}
\end{equation}
with the objective function being the intercluster pair distribution. We
have performed these calculations for the aforementioned
densities. With the temperature defined as  k$_B$T/$\zeta$ in terms of the potential well
depth, ($\zeta = 0.068$ kcal/mol, see Figure \ref{ur}), 
The cluster-cluster
potentials have been evaluated for the lowest temperature considered,
k$_B$T/$\zeta = 3$. The double volume integrals over the cluster volumes are carried out
numerically using a Gauss-Legendre quadrature, with $u(r)$ given by
Eqs.~(\ref{SALR}) and (\ref{pot}). With this procedure we end up with
estimated effective cluster densities of  $\rho_{cl}^{eff}\sigma^3=1.2$
($\rho\sigma^3=0.05$) and 0.83 ($\rho\sigma^3=0.025$), with
corresponding effective radii, $R_{cl}^{eff}/\sigma= 4.5$, and 4, respectively. 
 The results obtained from Eq.~(\ref{uav}) are presented
in the right graph of Figure \ref{ur}. The dashed curves correspond to the
cluster-particle interaction that is simply obtained by removing one of
the effective cluster densities and one volume integration from
Eq.~(\ref{uav}). We see in the Figure that the net
inter-cluster interaction is extremely repulsive, orders of magnitude
more intense than the interparticle potential, and dies out
at $\sim \lambda_0$. This explains why this quantity is
closely connected with  the intercluster separation, as 
reflected by the first peak of the cluster-cluster pair distribution
that will be discussed in the next section.   Our spherical clusters
act like large repulsive spheres of an approximate exclusion diameter $\lambda_0$,
whereas their actual diameter (the precise value depends on the definition
as we will see) is $\approx \lambda_0/2$, i.e. the excluded volume is
2$^3$ times the average cluster volume. This is illustrated in Figure
\ref{clus-conf} where for a simulation snapshot the excluded volume is
represented by a transparent sphere. The effective cluster density as
estimated from the excluded volume is in this instance
$\rho_C\lambda_0^3 = 0.82$, where $\rho_C$ is the number density
of clusters. Note that this effective density corresponds to a rather
dense fluid, in particular when compared with the very low net
particle density ($\rho\sigma^3=0.05$).

Molecular dynamics simulations have been  run specifically in the NVT ensemble using the LAMMPS
package\cite{Thompson2021} for a total of 218000 particles of mass 100
 amu. Distance
and energy units for the simulation have been chosen following the 
LAMMPS package {\it real} units specification. The integration time step was set to 2
fs. Results will be presented in terms of reduced units, both for
temperature (defined in terms of the well depth as indicated above),
length (using $\sigma=4$\AA\  as unit length) and time, for which the
reduced time unit is given by
$\tau_0=(m\sigma^2/\zeta)^{1/2}=7.5$ps. The systems were equilibrated for 20
ns, of which the first 10 ns 
correspond to a  slow cooling process following a linear temperature ramp starting from a fully
disordered system at k$_B$T/$\zeta = 18$ down to the desired
temperatures, namely, k$_B$T/$\zeta$ = 3, 6, 9, 12, and 15 for $\rho\sigma^3= 0.05$ and
just  k$_B$T/$\zeta$ = 3, and 9 for $\rho\sigma^3= 0.025$. It is worth stressing that
shorter equilibration runs (5 ns) lead to very similar cluster size/radii
distributions. This is an indication that the results presented
correspond to equilibrium states. For much shorter equilibration runs
and/or cooling times it is very easy to end up with bimodal or
even multi-modal cluster size/radii distributions.  Production runs were
$\tau_{run} = 20$ns long with configurations stored every 10 ps.

 \begin{figure}[t]
\centering
\includegraphics[width=12cm,clip]{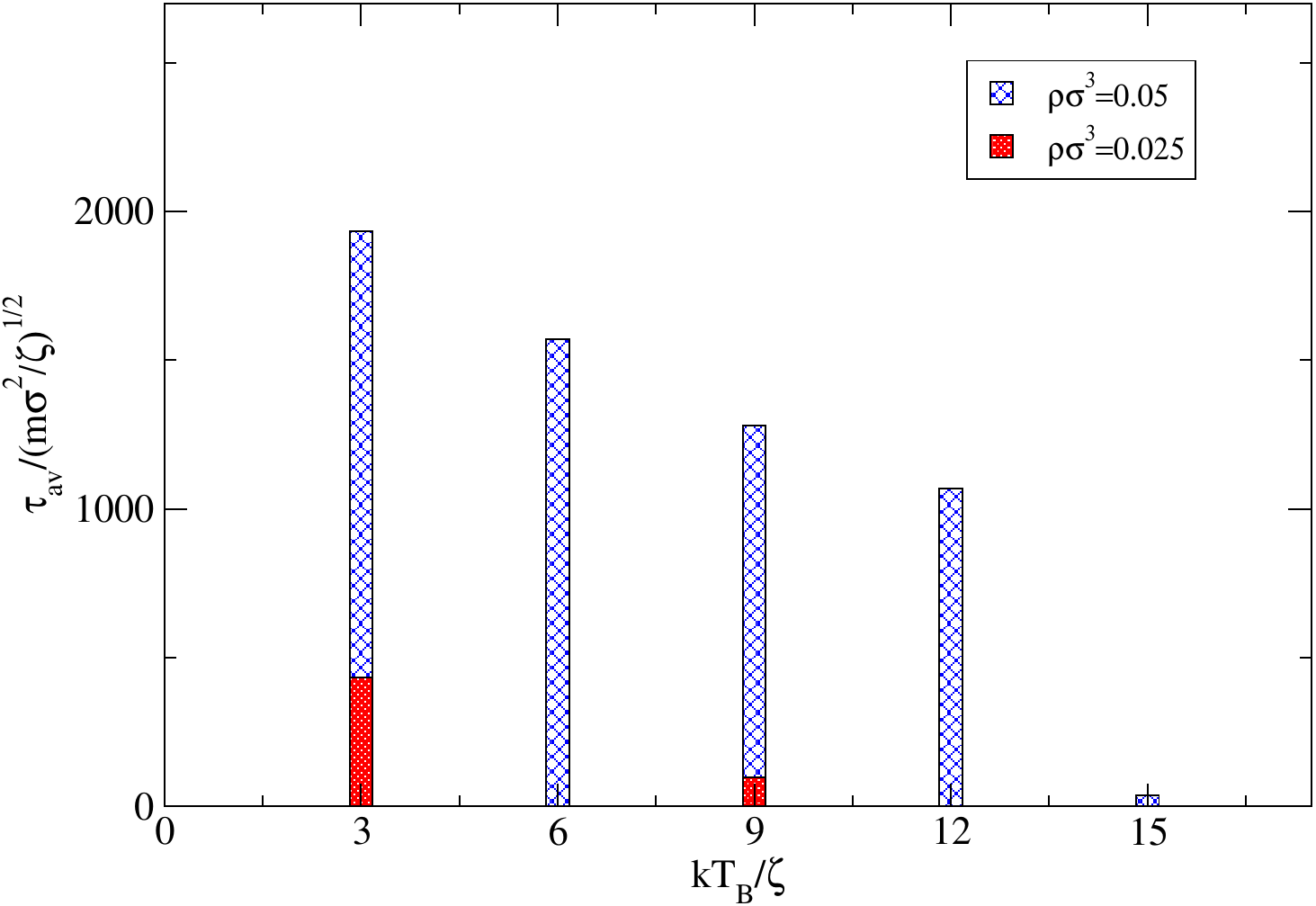}
   \caption{Average life time of the clusters, ($\tau_{av}$,
     as a function of temperature for $\rho\sigma^3=0.05$ (blue bars)
     and $\rho\sigma^3=0.025$ (red bars). The upper
     limit of the $y$-axis corresponds to the length of the simulation
     production  run, $\tau_{run}$. Note that for
     $\rho\sigma^3=0.025$ the only temperatures calculated are
     $k_BT/\zeta = 3,$ and 9. \label{clus-stat}}
\end{figure}

\begin{figure*}[t]
\centering
\subfigure[]{\includegraphics[width=8.5cm,clip]{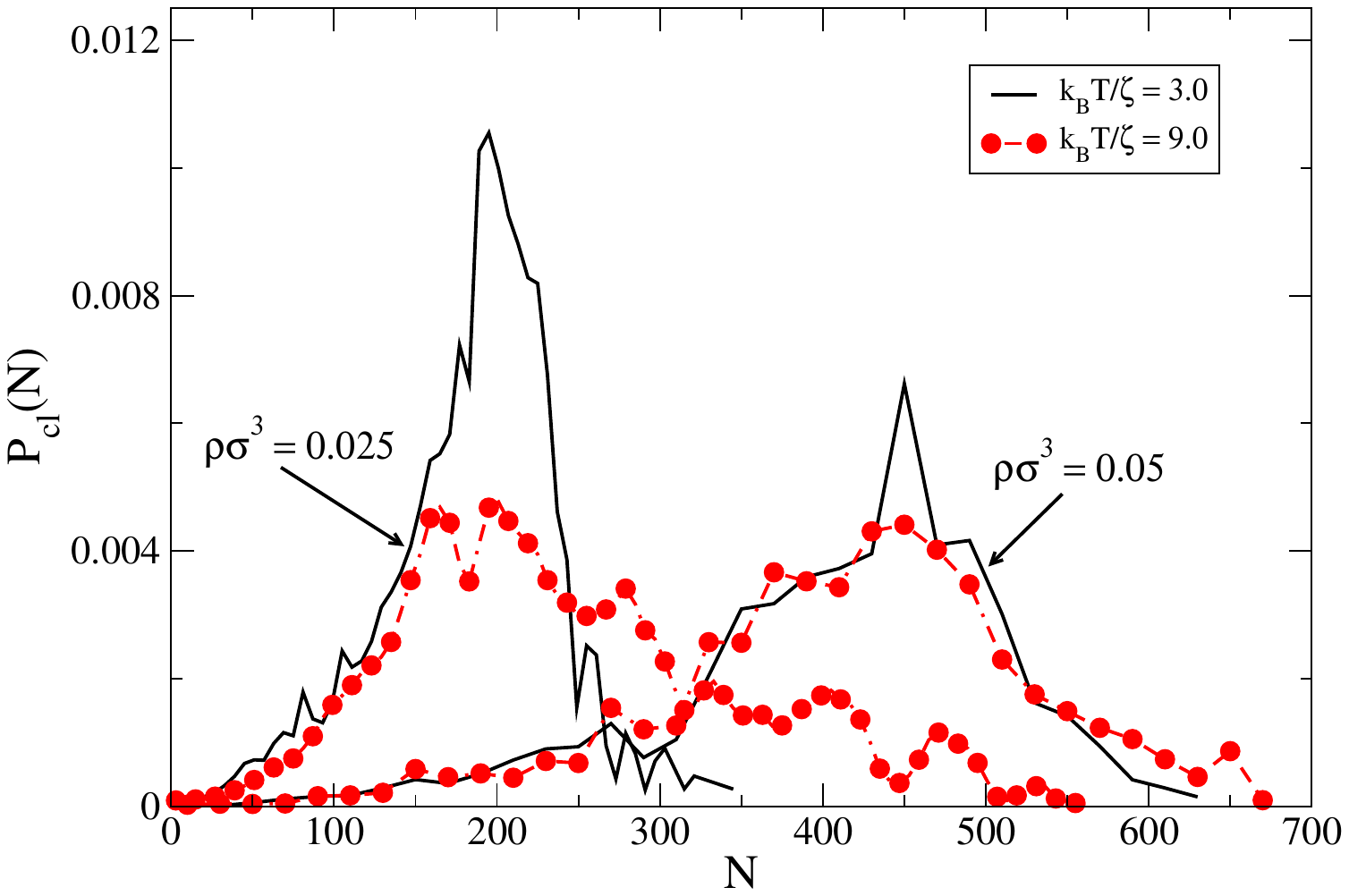}\label{cldis}}
\subfigure[]{\includegraphics[width=7.5cm,clip]{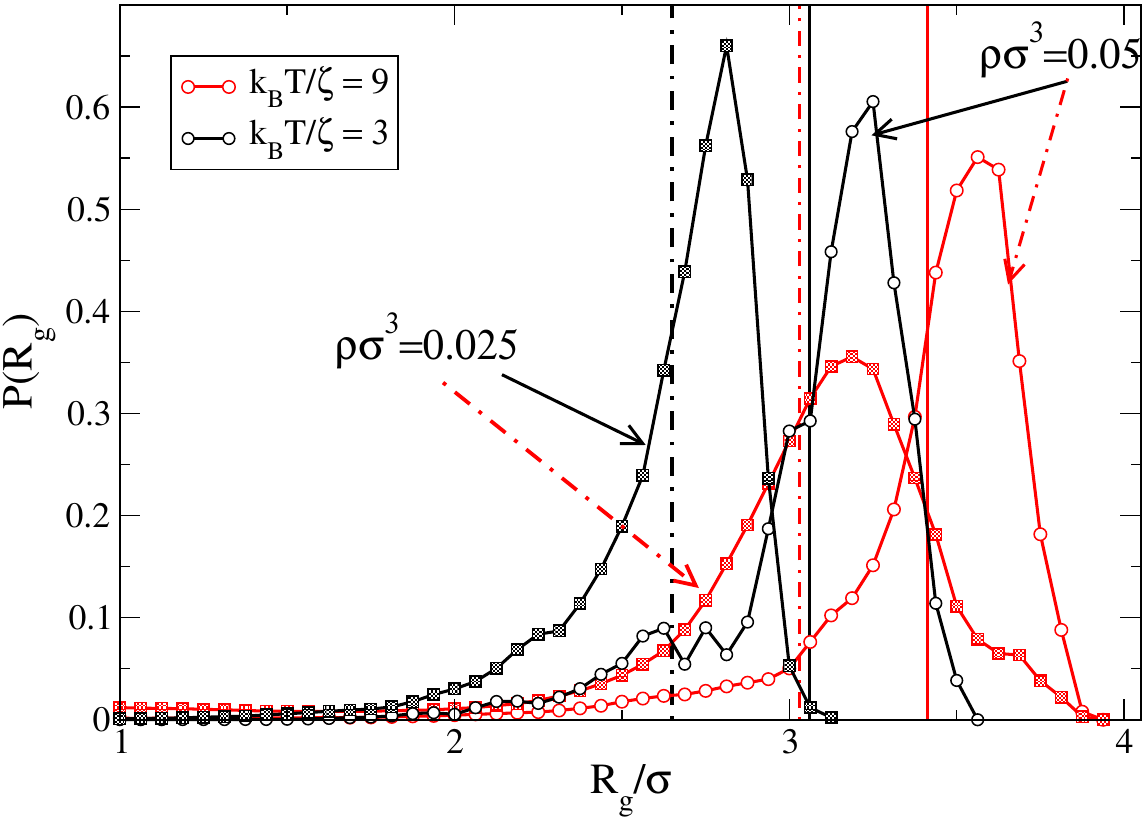}\label{diamf}}
   \caption{a) Cluster size distribution dependence on temperature and
     global density b) Cluster radii of gyration distribution in terms of
     density and temperature. Average radii are indicated by
     vertical lines, with solid ones corresponding to
     $\rho\sigma^3=0.05$ and dash-dotted ones to $\rho\sigma^3=0.025$.\label{clusan}}
\end{figure*}

\section{Onset of clustering\label{onset}}
As mentioned before, we have chosen potential parameters and
thermodynamic conditions for the system to yield a globular cluster
phase. We have first explored a series of temperatures for the
two densities indicated above using the Hypernetted Chain integral
equation \cite{HansenBook3rd}. As shown by Archer and
Wilding\cite{Archer2007a} this equation has a non-solution region
which hides the cluster phase. This lack of solutions (in fact
solutions have analytical continuation in complex space\cite{Lomba1995}) can be
interpreted as a signature of a phase transition,  in this instance a
first order transition between the particle gas and a fluid of liquid-like
spherical clusters. For a very closely related system, this first order transition was
identified by Archer and Wilding\cite{Archer2007a} using both Density
Functional Theory and extensive Grand Canonical Monte Carlo simulations
exploiting multi-canonical reweighting aided by  multi-histogram
reweighting techniques\cite{Archer2007a}. In our case we find that real solutions for the largest
density disappear below $k_BT_{ns}^{HNC}/\zeta =
14.9$, and $k_BT/\zeta = 8$ for $\rho\sigma^3=0.025$. For
$k_BT/\zeta = 17.5$  and $\rho\sigma^3=0.05$ the HNC converges yielding
a structure factor in excellent agreement with the simulation (cf. Figure \ref{sqhnc}).
A visual inspection of simulation snapshots does not show  evidence  
of clustering. Interestingly,  in Figure 
\ref{sqhnc} we first see that  there is a
significantly high prepeak (also
known as intermediate range order peak \cite{Godfrin2014}) which reaches the value
$S(Q_{max})\approx 2.7$ for $Q_{max} \approx 2\pi/\lambda_0$, whereas 
 for $Q\sigma \gtrsim 1$ the structure factor corresponds to that of a
 dilute gas of uncorrelated particles (cf. the inset in Figure \ref{sqhnc}).  Godfrin and coworkers propose
 the use of this precise value of the prepeak height as signature
of the onset of the cluster phase\cite{Godfrin2014}. We will see in
the next sections that lowering the temperature increases the height
of the prepeak by orders of magnitude, i.e. $S(Q_{max})$ practically diverges for a
$Q_{max}\sim Q_0 \ne 0$, which suggests the existence of a non-zero Lifshitz
point\cite{Bomont2012a}. This means that the system becomes unstable
with respect 
to a modulated inhomogeneous phase, whose modulation is given by
$Q_{max}\approx 2\pi/\lambda_0$. The locus of Lifshitz points for varying
densities yields the $\lambda$-line \cite{Archer2007a}, a term
borrowed from the study of criticality in ionic
fluids\cite{Stell1995}. It must be mentioned, that Bollinger and
coworkers\cite{Bollinger2016} suggested the use of an additional
criterion to characterize the onset of the cluster phase, namely, 
a thermal correlation length with values within the interval $2\lesssim
\xi_T/\sigma \lesssim 3$. This quantity can be estimated from 
a Lorentzian fit
of the prepeak, which in our case yields  $\xi_T/\sigma = 5.6$,
exceeding the limit suggested in Ref. \cite{Bollinger2016}. 
However, one must bear in mind
that  Ref. \cite{Bollinger2016} is focused on Yukawa (screened
Coulomb) interactions, 
and here we are dealing with bare exponentials. So far it has not been
assessed to what extent the choice of the explicit form of the
potential affects the reliability of this criterion. Until a complete
analysis of the type performed in Ref.~\cite{Bollinger2016} is
carried out it seems reasonable to stick to Godfrin's et al
choice.

An illustration of a globular cluster configuration can be seen in
Figure \ref{clus-conf}, where we have also represented the effective
excluded volume of the clusters using transparent spheres. One
can see that despite a relatively low concentration of clusters,
the spheres representing the excluded volume are arranged in a rather
dense packing, which will be reflected both in the microscopic
structure and dynamics of our system.

\section{Single-cluster structure and dynamics\label{oneb}}

In order to define the clusters we have used a geometric criterion,
namely, all those particles whose separation is below $d_{cl} = 1.3\sigma$ are
considered as pertaining to the same cluster. This value is obtained
from a preliminary run at the lowest temperature and
$\rho\sigma^3=0.05$ for which the particle-particle distribution
function, $g(r)$, was evaluated. We found that  $d_{cl} = 1.3\sigma$ 
roughly corresponds to the first non-zero minimum of $g(r)$, i.e. it is an estimate
of the outer boundary of the first coordination shell. Once this
clustering distance was set, we have run a GPU density based parallel scan
(G-DBSCAN  \cite{Andrade2013}) using an in-house code. Specific GPU
optimizations were performed to calculate structure factors, pair
distribution functions \cite{Levine2011,Sakharnykh2015,Reuter2019} and
dynamic properties. Only clusters containing more than 4 particles
have been considered for the analysis below.

A first observation from our cluster analysis is the fact that the
structures are long-lived. In Figure \ref{clus-stat} the average
cluster life time, $\tau_{av}$ is displayed for the thermodynamics
states under consideration. This quantity is defined as the average of
the time a given cluster retains its identity (does not merge, split
or dissolve, but it might incorporate/lose individual particles) in the course of the
simulation. The top of the ordinate axis 
corresponds to the length of the production run, $\tau_{run}\approx
2700\tau_0$. For the highest density, up until $k_BT/\zeta = 15$ the
average lifetime exceeds 
half the length of the simulation, which is an indication of the
stability of the structures. This is explained by the large depth of
the cluster-particle potential (see right graph of Figure
\ref{ur}). At  $k_BT_c/\zeta \approx 15$ the clusters are extremely
short lived, this temperature being fairly close to the limit of real solutions
of the Hypernetted Chain integral equation, which has been discussed
above. For higher temperatures no significant clustering 
is found. Now, when the density decreases the
average life time of the clusters decreases substantially. The real
solution of the HNC disappears at $k_BT/\zeta = 7.6$, but for
$k_bT/\zeta =9$ the simulation results still display a significant
degree of clustering, although much more short lived. Simulation runs
for higher temperatures (results not included) do not show evidence of
clusters satisfying our minimum size criterion. In Figures S1 in the
Supplementary Information one can see snapshots of the evolution of the cluster
formation as the system is cooled down from  $k_BT/\zeta = 18$ to  $k_BT/\zeta = 3$.

 \begin{figure}[b]
\centering
\includegraphics[width=12cm,clip]{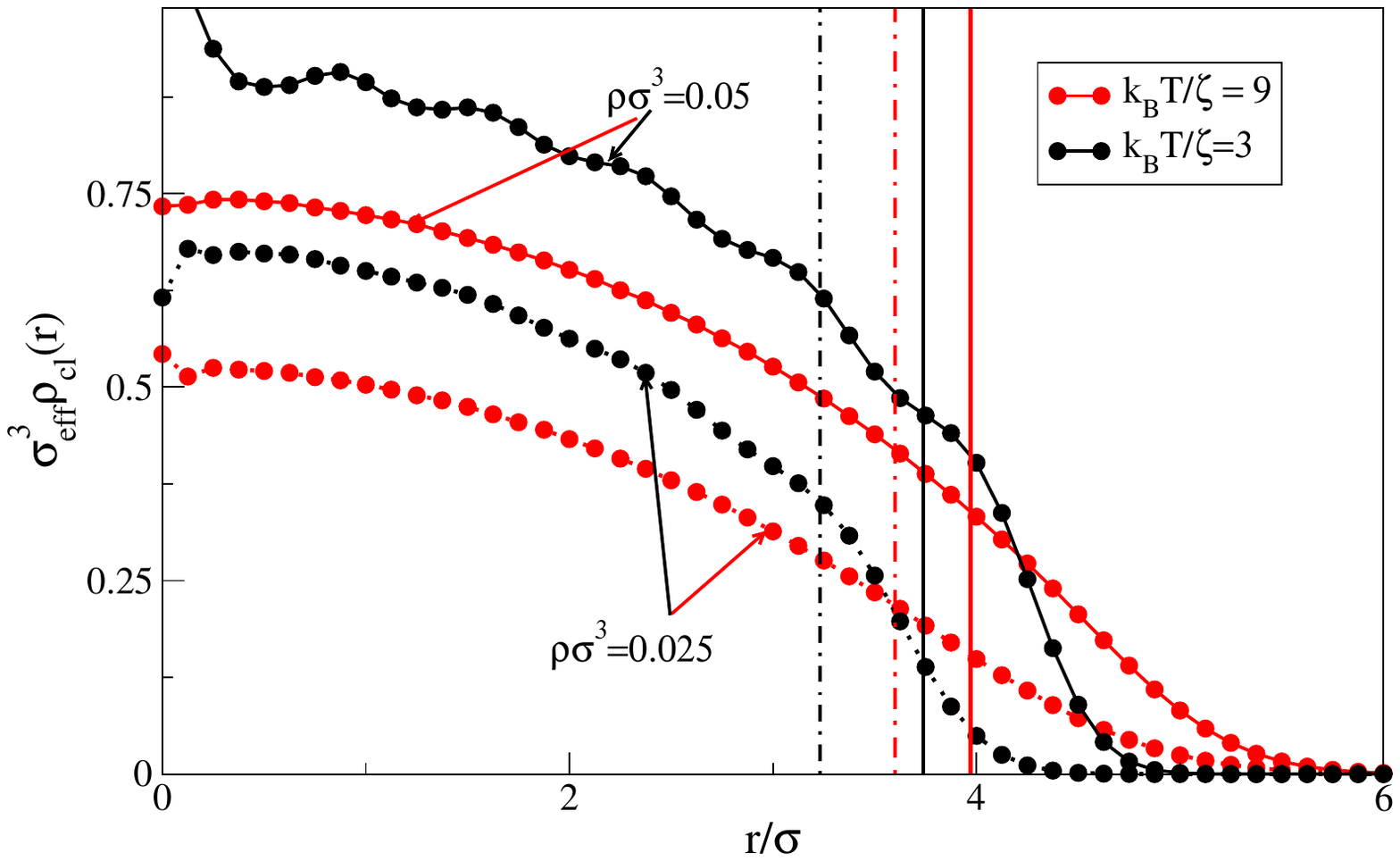}
   \caption{Temperature and density dependence of the average density
     profile of the clusters. Densities have been scaled with the
     effective diameter of the particles which is
     $\sigma_{eff}\approx 0.8\sigma$. Vertical lines denote the radii
     corresponding to the Gibbs
     dividing surface (cf. Eq.~(\ref{RG}) in the text),  with solid ones corresponding to
     $\rho\sigma^3=0.05$ and dash-dotted ones to $\rho\sigma^3=0.025$.\label{rhoclus}}
\end{figure}

 \begin{figure}[t]
\centering
\includegraphics[width=12cm,clip]{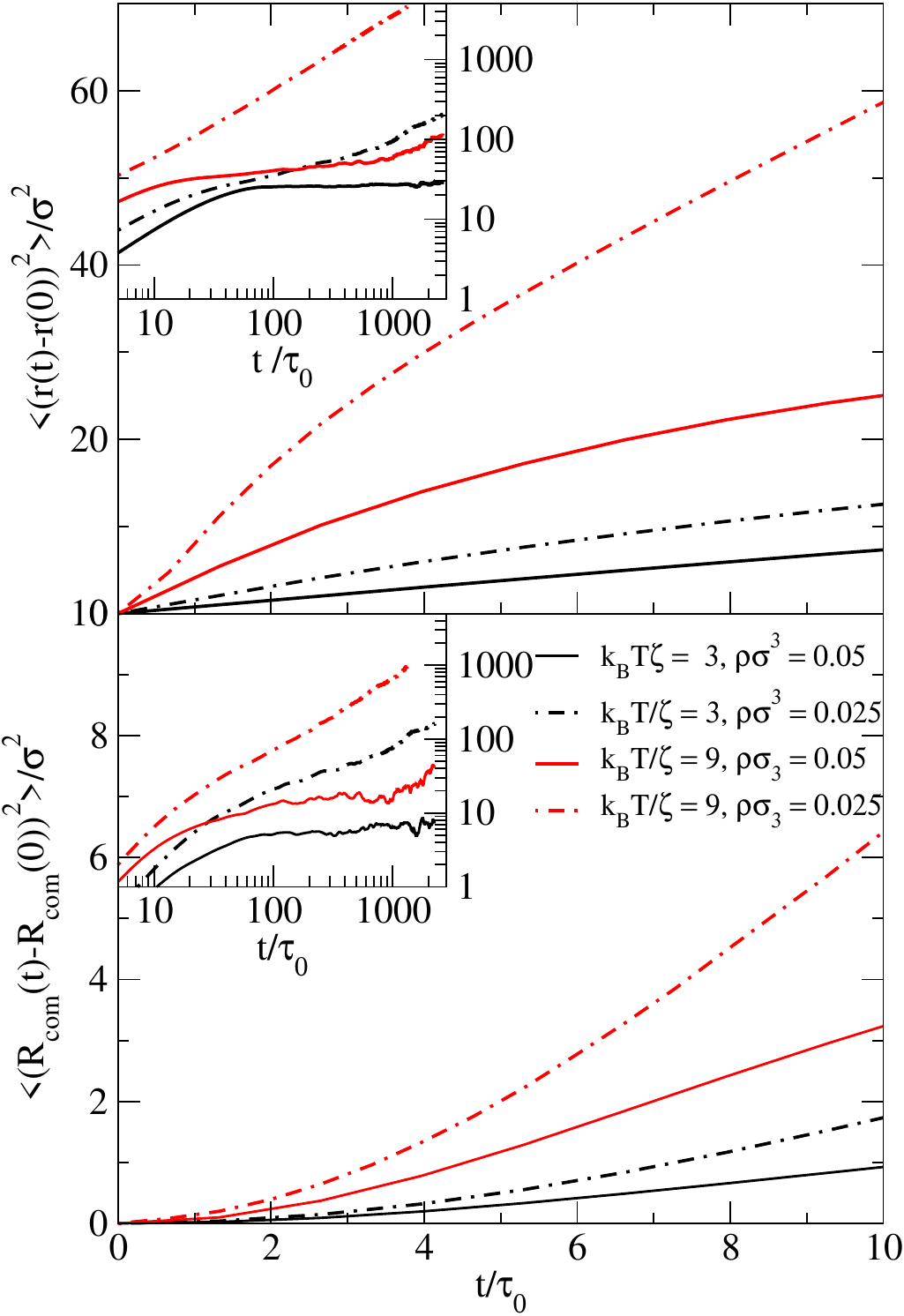}
   \caption{(Upper graph) Single particle mean square displacement dependence on
     temperature and density. (Lower graph) Density and temperature
     dependence of the center of mass mean square
     displacement for persistent clusters. In both graphs velocities
     and time are expressed in reduced units. In the insets the long
     time behavior is illustrated using log-log plots.\label{diffu}}
\end{figure}

As to the cluster statistics, in Figure \ref{clusan} we present the
cluster size distribution for the two densities in question and two
selected temperatures (left graph) and the corresponding cluster
radii distributions depicted in the right graph. As estimate of the
cluster radii we have 
used  the value of the gyration radius, which is defined in the
usual fashion \cite{IUPAC2006} by
\begin{equation}
R_g = \left<\sqrt{\frac{\sum_i m_i ({\bf r}_i-{\bf R}_{com})^2}{\sum_i
      m_i}}\right>
\label{Rg}
\end{equation}
where the brackets indicate a ensemble average and ${\bf R}_{com}$ is
the center of mass of the cluster whose radius of gyration is being
computed.  We see that for the largest
density the size distribution peaks at around 450 particles and at 200
particles for the lowest. For all temperatures the clusters are liquid
droplets, as will be shown below when analyzing the short time
behavior of the mean square displacement. 

Interestingly one can see that the shape of 
the cluster size distribution is hardly affected by temperature in the case of the
largest density, 
whereas for low density it becomes much wider and much larger clusters
appear as temperature rises. This, in combination with the short
lifetime of the clusters seen in Figure \ref{clus-stat}, is an
indication of a rapid merging and splitting of clusters taking place
for the lower density/high T case. The large spread of the low density-high T
size distribution is also reflected in the cluster radii
distribution, which in this case is almost Gaussian-like. All
remaining distributions are 
negatively skewed beta-distributions with a more or less wide hump at
approximately one particle diameter below the maximum of
$P(R_g)$. Large clusters are in equilibrium with smaller ones,
coalescing with them and letting particles escape their outer
atmosphere to subsequently form  smaller clusters. Merging of larger clusters
is prevented by the large intercluster repulsion (cf. see right graph
of Figure
\ref{ur}), and this is at the root of the beta-distribution shapes.  For the low
density/high temperature case thermal energy is sufficient to overcome
the intercluster repulsion to a certain extent, and the equilibrium between destruction and creation of
clusters leads to a Gaussian-like distribution. 

This picture is complemented by an inspection of
the cluster average density profiles, $\rho_{cl}(r)$, depicted in Figure
\ref{rhoclus}. In this case, we have scaled the densities with
$\sigma_{eff}=0.8\sigma$ in order to be able to compare the densities
with those of standard fluids (e.g. for a  Lennard-Jones fluid, liquid
densities are usually assumed to be $\rho\sigma^3\gtrsim 0.4$).
One first notices that despite the extremely 
low total particle densities, for $\rho\sigma^3=0.05$ the clusters 
display very high values of the effective density profile, even for
high T. These values are  well into the
liquid domain. For the lower
global density, the effective inner cluster density remains practically within the
boundaries of the liquid state. In both cases there is a transition
towards gas-like densities as the profile approaches the surface in a
gradual fashion. Taking into account that the clusters are
rather spherical, we have defined a  Gibbs dividing surface
  in order to have a
quantitative description of the separation between gas-like particles and
liquid-like particles, as is customary in the description of gas-liquid
interfaces \cite{HansenBook3rd}. This surface places an ideal separation between the
gas-like cluster region from its corresponding liquid-like
counterpart. Here, for a spherical cluster this should be the surface
of a sphere, whose radius (Gibbs dividing radius) $R_{Gibbs}$ is given by
\begin{equation}
\int_0^{R_{Gibbs}} r^2(\rho_{cl}(0)-\rho_{cl}(r))dr = \int_{R_{Gibbs}}^\infty
r^2\rho_{cl}(r)dr.
\label{RG}
\end{equation}
Obviously, when one has a uniform density, $\rho_{cl}(r)=\rho_{cl}(0)$
with a sharp interface at 
$R_{Gibbs}$, both sides of Eq.~(\ref{RG}) are identically zero.  

We see that these Gibbs radii separating gas-like and liquid-like
particles in the clusters (vertical lines in Figure \ref{rhoclus})
follow the same trends as the average radii of 
gyration (vertical lines in Figure \ref{diamf}), i.e. larger values
for larger densities, and for the same 
density the radius increases with temperature. This latter trend reflects the
larger spread of the cluster size distributions as temperature
raises. The Gibbs radii are systematically larger ($\approx 0.5\sigma$)
than the gyration radii. This is a consequence of the lesser weight of
the ``gas atmosphere'' surrounding the quasi-spherical clusters in
Eq.~(\ref{Rg}). It is worth noticing that the cluster density profiles
reach vanishing values in the interval $4.5\sigma<r <5.5 \sigma$, by
which the average cluster spatial extent is $\approx \lambda_0/2$, in
accordance with the estimated obtained from the potential's
characteristic wave vector (cf. Figure \ref{ur}). The
temperature and density dependence of both $R_{Gibbs}$ and $R_g$ found
in this paper agrees with the finding of Schwanzer et
al.~\cite{Schwanzer2016} for much smaller clusters in two
dimensions. It is important to stress that in our system further increases of
density maintaining the temperature fixed would lead to an
increase in the number and size of the 
clusters, that would eventually coalesce to form lamellar phases which
finally  mutate into a bubble phase. Other SALR interactions (or different potential
parameters) at sufficiently low temperatures can
yield a wide panoply of modulated structures, such as  cluster crystals, gyroid
phases\cite{Edelmann2016}, crystals of bars, etc.   

 \begin{figure}[t]
\centering
\includegraphics[width=12cm,clip]{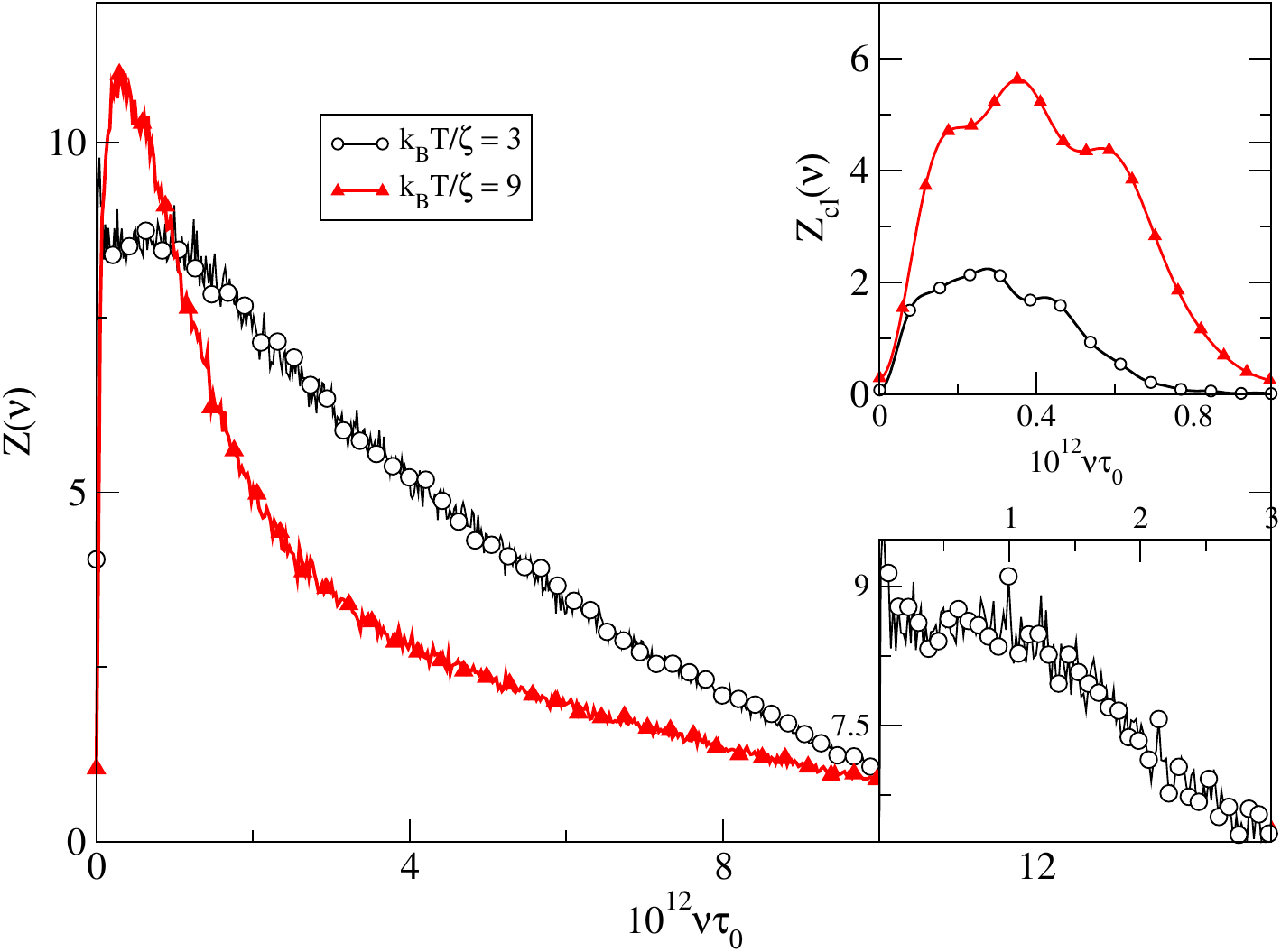}
   \caption{Vibrational frequency spectra derived from the particles
     velocity self-correlation at various temperatures and
     $\rho\sigma^3=0.05$. A detail of the low frequency region is
     presented in the inset in the lower right corner.  The cluster vibrational frequency
     spectrum as computed from the velocity
     self-correlation 
     of the persistent clusters centers of mass is depicted in the
     inset in the upper right corner.\label{zw}}
\end{figure}

 \begin{figure}[t]
\centering
\includegraphics[width=12cm,clip]{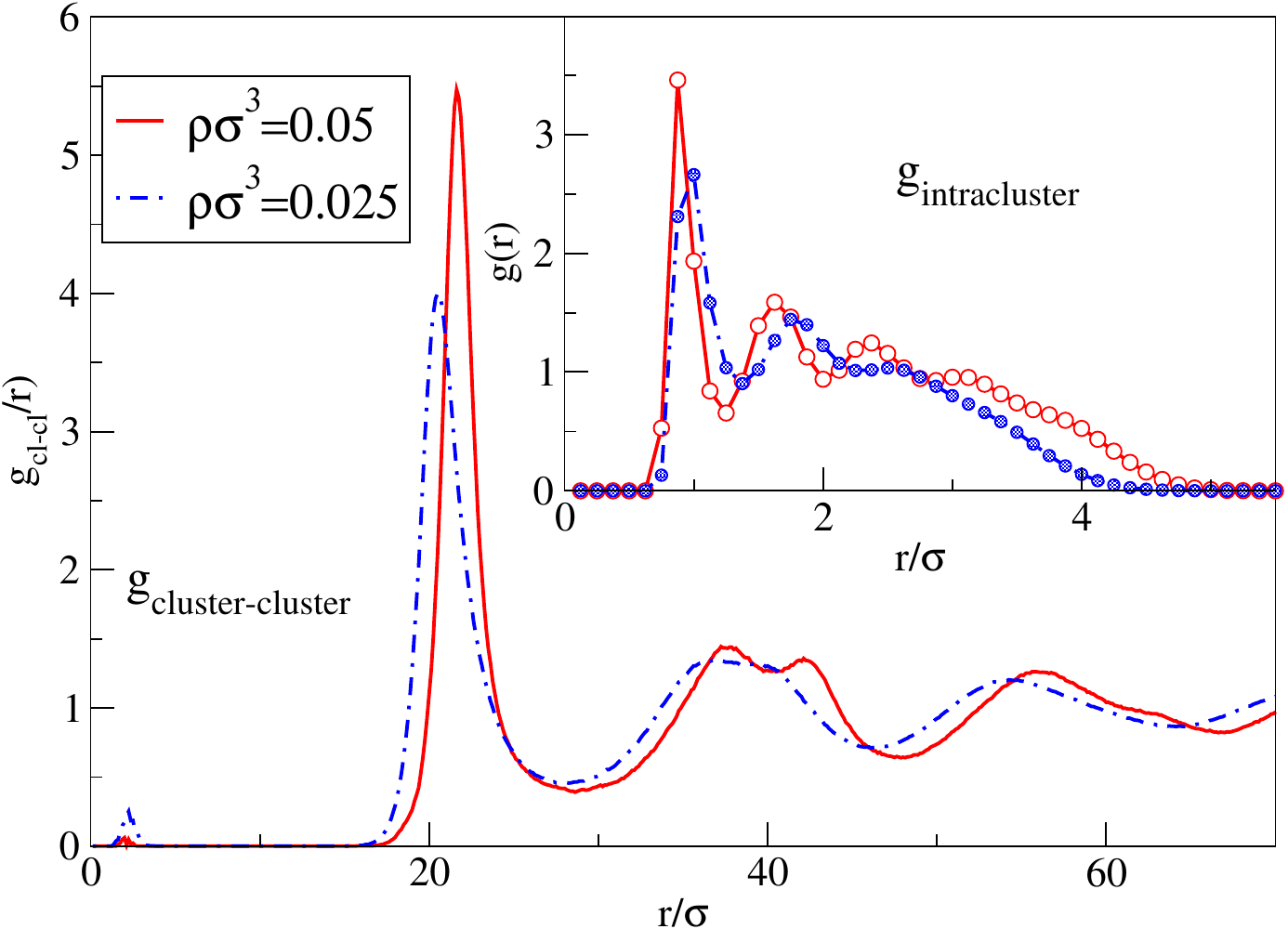}
   \caption{Cluster-cluster pair distribution function and
     intra-cluster pair distribution function (inset) for the lowest
     temperature state $k_BT/\zeta = 3$.\label{gtot}}
\end{figure}

Now a few words concerning the single particle dynamics. In what
follows,  in order to make the results independent of specific details
of the model, time and frequency will be reduced with the corresponding
molecular time unit, $\tau_0$ and its inverse
$\tau_0^{-1}$ respectively. Here we first
look at the individual particle mean square displacement (m.s.d), plotted in
the upper graph of Figure \ref{diffu}. It
is apparent from the figure that there are clearly several separate
regimes. First, there is a short time scale (under 2 $\tau_0$, i.e. for our system around
14ps) where the system is in a diffusive regime corresponding to 
intracluster movements (the mean square displacement is $\lesssim
16\sigma^2\approx <R_g>^2$). Then the system reaches a subdiffusive regime with a decreasing diffusion constant (slope of the
curves). This reflects the finite size of the clusters. For much
longer times (up to 2.7$\times 10^3 \tau_0$) the behavior of the mean
square displacement is represented in the inset in log-log scale. The
low density-high temperature case displays a fully linear behavior in
the m.s.d. In this
  case the regime is consistently diffusive at all times, except for
  the short lived ballistic region at very short times, hardly visible in
  the graph. In contrast, the two systems
with the largest density  display a plateau
after the initial diffusive and subdiffusive regimes, which is then
followed by a long time slightly diffusive
behavior, hardly visible in the case of low T and highest density.  This
long time behavior stems  from the collective cluster movement
which is shown in the lower graph of Figure \ref{diffu}. Here we
have plotted the mean square displacement of the center of mass of
persistent clusters, that is, those that preserve their identity
during the length of the production run. At short times one can even
appreciate the ballistic regime with a parabolic m.s.d. before the
clusters begin to collide. We see that the long time behavior
(inset in log-log scale) displays the same trends as the single
particle mean square 
displacement.  One can actually measure a tiny diffusivity for the
clusters at higher density/low temperature, namely
$D/(\sigma^2\zeta/m)^{1/2} = 0.0003$. If this quantity is rescaled to
take into account the average mass of the clusters we would still have
a very small diffusivity, $D/(\sigma^2\zeta/M)^{1/2} \approx 0.006$.  
This is an indication that our
system is approaching an arrested glassy state, and the dynamic slow
down is not only due to the larger mass of the clusters. Taking into account
that the source of this quasi-freezing of the cluster positions is
mostly induced by the long range of the repulsive interactions (cf
the right graph of Fig.~\ref{ur}) one is tempted to identify this
state as a precursor of a Wigner glass of 
clusters\cite{RuizFranco2021}. However, strictly speaking the
formation of Wigner
crystals  and glasses \cite{Duang2005} is due to extremely long ranged
interactions such a Coulombic ones, in conjunction with higher packing
fractions. Following the discussion of Klix 
and coworkers \cite{Klix2010} and bearing in mind the significant
interparticle attraction and low particle packing fraction of our model, we are
most likely somewhere near the 
dynamic transition towards a cluster glassy state. It is worth
  recalling that despite the very low particle density, as mentioned
  before, (cf Figure \ref{clus-conf}) due to the large
intercluster repulsion the effective density as inferred from the
cluster excluded volume lies in the range $\rho_C\lambda_0^3 = 0.74
\sim 0.82$ for all cases studied here, corresponding to relatively
dense fluid states. 

These features are further confirmed by an analysis of the velocity
self-correlation functions which are illustrated and commented upon
in the Supplementary 
Information (cf. Figure S2).  These quantities, once Fourier
transformed provide the 
frequency spectra, which are plotted in 
Figure \ref{zw}. We see first that  the high values of 
$Z(0)$ confirm the 
liquid-like dynamics of the cluster constituent particles. For
$k_BT/\zeta =3$ and $\rho\sigma^3=0.05$,  cage effect vibrations
are reflected as a tiny maximum (see lower inset of Figure \ref{zw})
observed at  
$10^{12}(m\sigma^2/\zeta)^{1/2}\nu\approx 0.65$, whereas as T is  
increased a much wider maximum occurs at lower frequencies
$10^{12}(m\sigma^2/\zeta)^{1/2}\nu\approx 0.25$. This maximum stems
from the coupled particle-cluster dynamics, an effect   also seen in the
velocity self-correlation function depicted in Figure S2. 
In the upper inset we have the corresponding  vibrational frequency
spectrum derived from the persistent cluster velocity self-correlation
function. Now the situation is completely different and the low
frequency behavior corresponds to a that of a system approaching a
glassy state, with very low diffusivities ($Z_{cl}(0)\longrightarrow
0$), in agreement with the 
long time behavior of the m.s.d. seen in Figure \ref{diffu}. Note
that the frequency domain of the cluster dynamics is one order of
magnitude smaller than that 
of single particles, reflecting the much slower dynamics of the
clusters, a consequence of  their  large masses. The intensity of the
peaks increases with temperature, as the kinetic energy increase
reflects mostly in 
vibrations with larger amplitude around equilibrium positions, with the maxima
shifting to slightly higher energies. This three maxima correspond to
three minima (cages) in the cluster velocity self-correlation function
(right graph in Figure S2), where also one can see that the cages
become narrower and move to shorter times as temperature is
increased. This effect is due to the higher vibrational energy
of thermal origin present in the high T state, which for
$\rho\sigma^3=0.05$ 
preserves some of the characteristics of a quasi-amorphous state. Lowering the
density and/or further increasing the temperature above
$k_BT/\zeta > 15$ obviously ends
up in the dissociation of the clusters and the complete disappearance of the
vibrational structure of the spectrum. 

 \begin{figure}[t]
\centering
\includegraphics[width=12cm,clip]{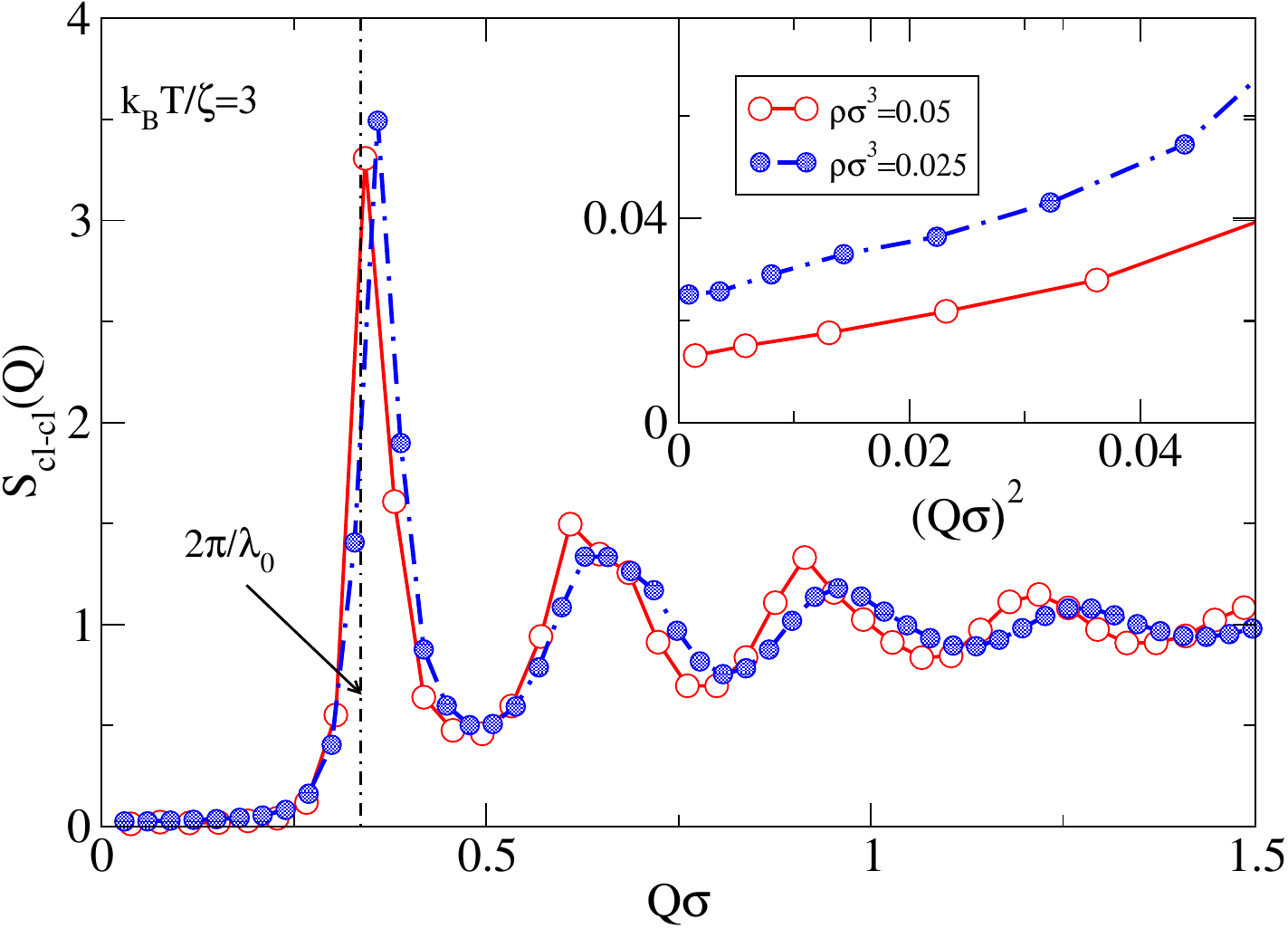}
   \caption{Cluster-cluster structure factor for the lowest
     temperature state, $k_BT/\zeta = 3$.\label{sqcl}. In inset the
     low-Q behavior is illustrated with using $Q^2$ as abscissa.}
\end{figure}

\section{Two-particle and intercluster structural
  correlations. Build-up of effective hyperuniformity\label{twob}}
\begin{figure}[t]
\centering
\includegraphics[width=12cm,clip]{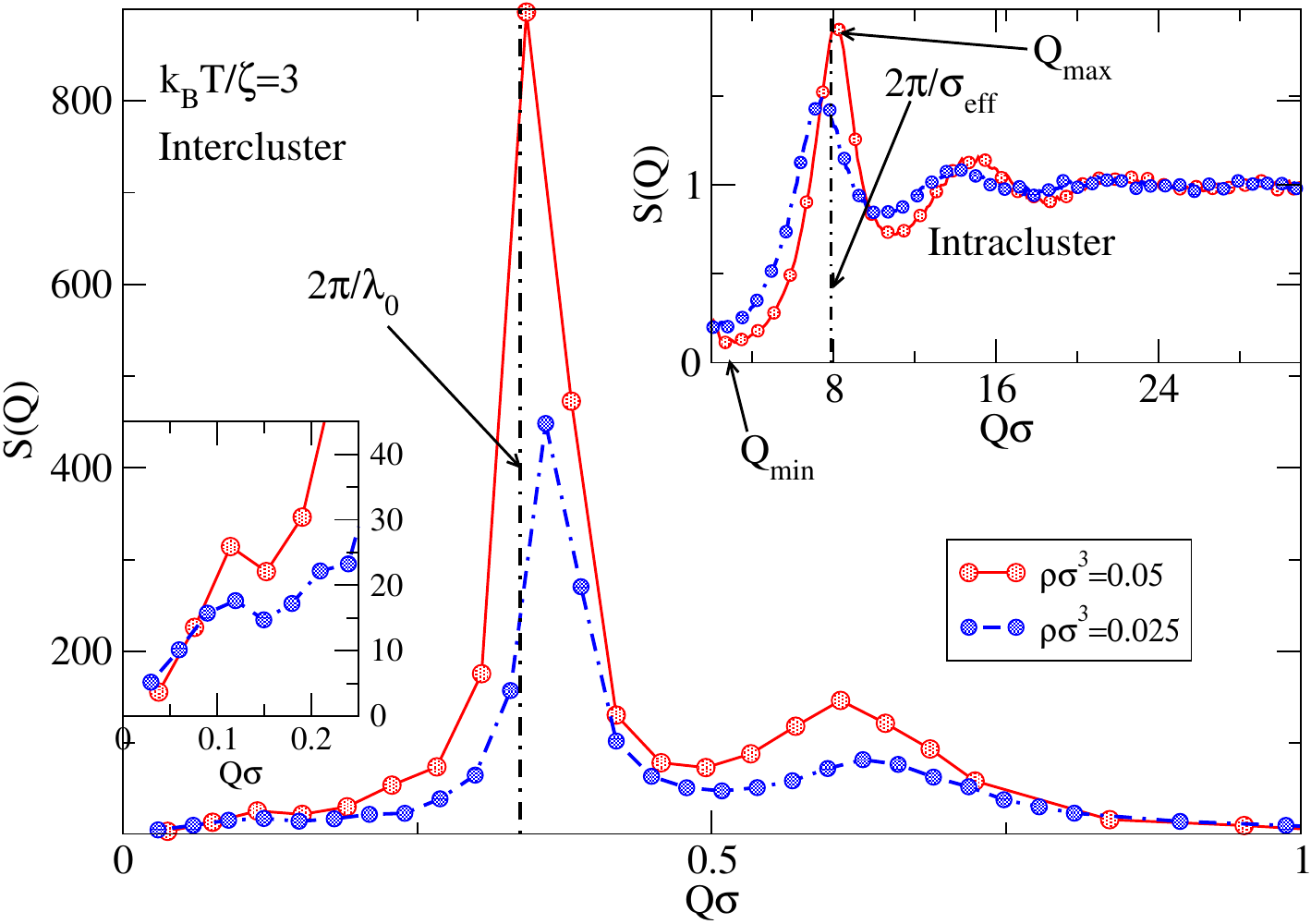}
   \caption{Density dependence of
     the total structure factor S(Q) (main graph) and the intracluster
     structure factor (upper right inset) for the system in stable
     cluster phases.  In the lower left inset the low-Q behavior of
     the particle-particle structure factor is depicted.\label{clus-sq}}
\end{figure}
We can now analyze two-particle correlations. To start with, in Figure
\ref{gtot} we present the cluster-cluster pair distribution function
for the lowest temperature and the two densities under
consideration. In the inset the corresponding 
particle-particle intracluster pair distribution functions are shown. Our 
previous discussion concerning the average intercluster potential
(cf. right graph in Figure \ref{ur}) explains that clusters are
separated by the correlation length $\lambda_0\approx 20\sigma$. The
smaller size of the clusters for low density 
implies the shift of the first peak of $g_{cl-cl}(r)$ to lower distances. The
splitting of the second coordination shell for $\rho\sigma^3=0.05$ is
a typical signature of the presence of an amorphous
structure\cite{Pan2011}, although in our case there is a residual
diffusivity.  This feature is found in dense amorphous
systems, but here the total number density is very low
($\rho\sigma^3=0.05$). However, we must bear in mind that the
excluded volume of the clusters is much larger than what would correspond
to their actual size, as previously discussed (cf. Figure
\ref{clus-conf}). Recall that the effective cluster diameter
(corresponding to the excluded volume) is
approximately twice the 
spatial extent of the clusters. On the other hand, once the particle density is
halved the splitting 
disappears and the system becomes more liquid-like, even if the
cluster diffusivity remains relatively low ($D/(\sigma^2\zeta/m)^{1/2}
= 0.006$). 

  As to the intracluster pair distribution function, its
decay beyond 3$\sigma$ is a consequence of the finite averaged cluster size (cf
the density profiles of Figure \ref{rhoclus}) but its liquid-like
structure is very remarkable, with a first coordination shell reaching
considerably high values. From the position of the first peaks and
their decay towards shorter interparticle separations, one can
appreciate that the potential core is relatively soft, and that is the
reason why we have considered an effective particle diameter
$\sigma_{eff}\approx 0.8\sigma$. A steeper repulsion in
Eq.~(\ref{SALR}) (e.g. a Lennard-Jones $r^{-12}$ term) would bring the
effective size closer to $\sigma$, but this would require a finer grid
in the potential interpolation used in the LAMMPS
package\cite{Thompson2021} which might lead to memory exhaustion in
the GPUs. 

If we now move from real to Fourier space, in  Figure  \ref{sqcl} we
have the cluster-cluster structure factor, $S_{cl-cl}(Q)$. A first feature is to be
noted, the height of the first peak for both densities is beyond
Hansen-Verlet's\cite{Hansen1969} freezing rule $S(Q_{max})\gtrsim 2.86$ for which a
transition towards a crystal is to be expected. As discussed above,
our system presents features of 
an amorphous solid, but with a somewhat peculiar structure factor. We see
that $S(Q) \approx 0$, $\forall \; Q \lesssim
0.2\sigma^{-1}$. In fact,  one finds that the hyperuniformity index
$H=S(Q_{max})/S(0< Q \lesssim 0.2\sigma^{-1})\approx 10^3$
is considerably high. While this value is still away from the
criterion for {\it near hyperuniformity}, $H\gtrsim 10^4$ proposed by
Atkinson and coworkers \cite{Atkinson2016} in maximally random jammed
systems, it represents a considerable attenuation of density
fluctuations. Actually it meets the criterion of {\it effective
  hyperuniformity}, $H\gtrsim 10^2$ proposed by Chen and
et al.\cite{Chen2018}. As to the low-$Q$ limit, it is illustrated in
the inset of Figure 
\ref{sqcl}. We appreciate the a linear dependence with $Q^2$.  Our
system, being ``effectively hyperuniform'', does not reach  the
limiting behavior  $\lim_{Q\rightarrow 0} S_{cl-cl}(Q) \approx 0$, but
we one gets closer as density increases (but this  is hardly affected
by temperature). This effect is most likely due to the increase of the
effective repulsion, since the internal cluster density and average
cluster size rise when the overall density is increased. The $Q^2$
dependence found in Figure 
\ref{sqcl} is consistent with
a class I (effectively) hyperuniform material \cite{Torquato2018a}. Local number
fluctuations will further confirm this result as shown below. 

Now a few words concerning the position of the first maximum of the
structure factor,  $Q_{max}$. We see that for $\rho\sigma^3=0.05$,  $Q_{max} \sim
2\pi/\lambda_0$ whereas  $Q_{max}(\rho\sigma^3=0.025) >
2\pi/\lambda_0$. This  reflects the smaller
size of the clusters for the lower density (see the corresponding
radii of gyration in Figure \ref{diamf}) and matches the shift to
lower $r$ of the first maximum of $g_{cl-cl}$ in Figure
\ref{gtot}. The quantity $\lambda_0$ is sort of a ``ground state'' estimate of the
cluster size, as temperature rises or density decreases its accuracy
is lower.

If we now look at the total structure factor, $S(Q)$ depicted in
Figure \ref{clus-sq}, its most noticeable feature is the huge magnitude of
the prepeak, which is the signature of a large degree of
clustering. Obviously the prepeak positions correspond to that of the first
maxima in the cluster-cluster structure factors discussed above. As to
the issue of particle hyperuniformity, the situation is less
clear. Even if the  hyperuniformity index meets the criterion of {\em effective
  hyperuniformity} as defined above, in the inset one can appreciate
that the small-Q values of S(Q) are far from small. The results for
$\rho\sigma^3=0.05$ seem to approach zero as $Q\longrightarrow 0$, but
our sample size does not allow for probing smaller $Q$-vectors. The
scattering from small clusters and free particles seems to destroy
hyperuniformity in a region where otherwise the cluster structure can be
deemed as effectively hyperuniform (compare the region for $Q\sigma \lesssim
0.2$ in Figures \ref{clus-sq} and \ref{sqcl}).

Now, in the inset we can see the large wavenumber behavior of S(Q). The
peak maxima in this region  $Q_{max}\sigma\approx 8.1$ reflects interparticle correlations at
$0.8\sigma$, which is precisely the effective particle diameter,
$\sigma_{eff}$. This is in accordance with the maxima of $g(r)$ in the inset of
Figure \ref{gtot}. Interestingly, $H=S(Q_{max})/S(Q_{min})=0.2\times
  10^2$ which is not very far from the criterion of effective
  hyperuniformity. Note that both $Q_{max}$ and $Q_{min}$
    correspond to the intracluster structure factor. A proper choice
  of the short range interparticle 
  interactions might actually drive the system closer to exhibiting an
  additional wavenumber range where  a considerable attenuation of
  radiation probes could be present.

\begin{figure}[t]
\centering
\includegraphics[width=12cm,clip]{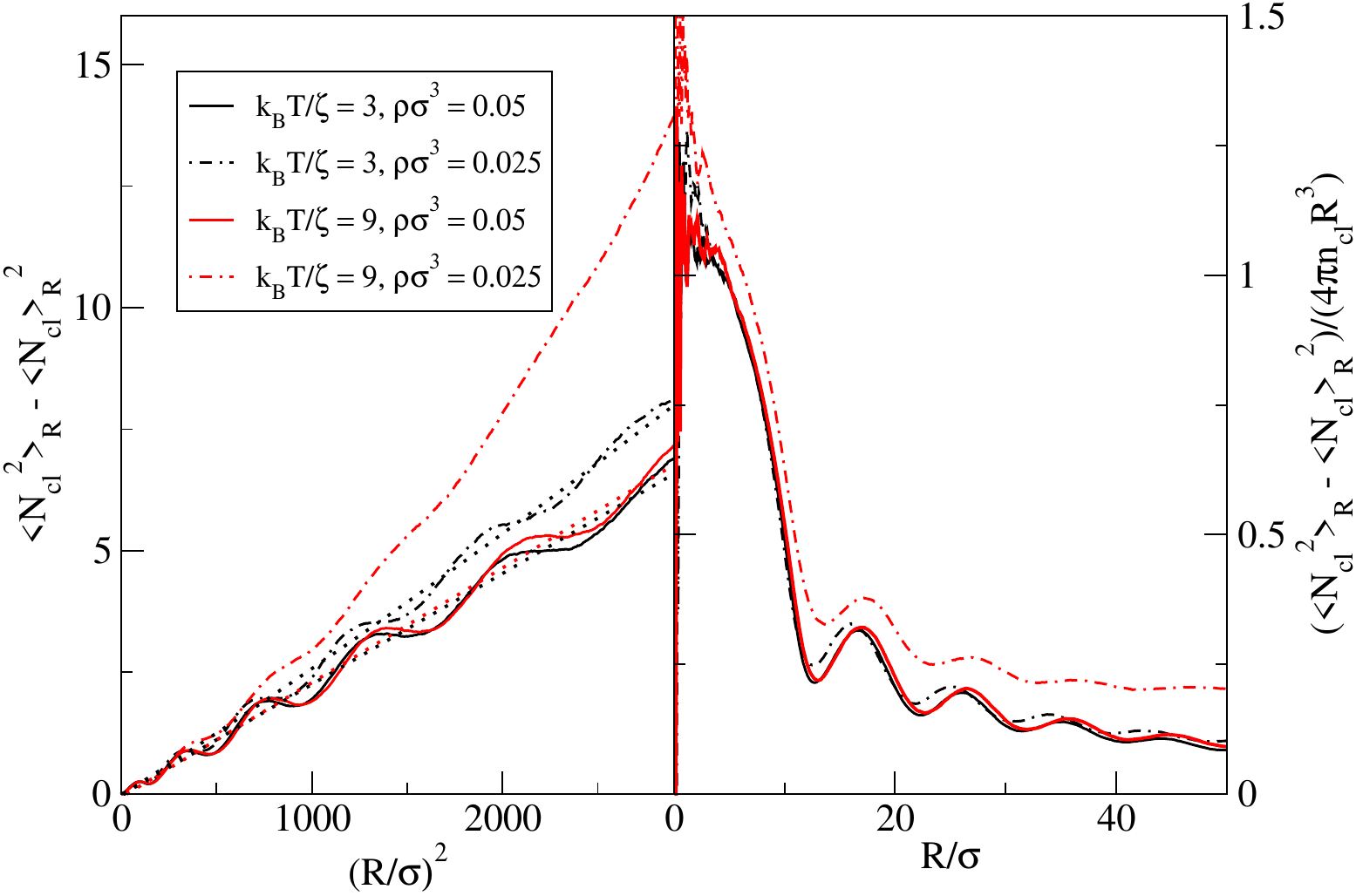}
   \caption{Left) Local particle number variation of cluster center of mass
     configurations vs the sample sphere radius squared, $R^2$. The linear
     dependence is an indication of a hyperuniform
     configuration. Deviation in linearity is apparent in the high
     T/low density case. Dotted lines represent linear regressions to
     the simulation data. Right)  Local particle number variation of cluster center of mass
     configurations scaled with the corresponding random uniform value
    , $4\pi n_{cl}R^3/3$ vs the sample radius, $R$. A decaying trend
    indicates the presence of a disordered hyperuniform system. Again,
    for the high
     T/low density case the scaled number fluctuation tends to remain
     constant. 
     \label{s2ncl}}
\end{figure}

As an additional assessment of  the effective hyperuniform character of
our systems, we have also analyzed the local density (or particle number)
fluctuations, defined as 
\begin{equation}
\sigma^2_N(R) = <N^2>_R - <N>_R^2
\label{s2}
\end{equation}
where $N$ refers to the number of particles and the subscript $R$
refers to the radius of  a sampling spherical volume. This
quantity is known to fulfill\cite{Torquato2003} $\sigma^2_N(R) =
<N>_R$ for a Poisson point pattern, which epitomizes a completely
random disordered and uncorrelated system. In fact, in many disordered
cases $\sigma^2_N(R) \propto R^d$, where $d$ is the dimensionality
($<N>_R\propto R^d$ in a uniform system). At the other end of the spectrum,  one
has point patterns known as {\em hyperuniform} that fulfill
$\sigma^2_N(R) \propto R^{d-\alpha}$ with $\alpha>0$. Among these  one
finds regular patterns 
such as those of crystals and quasicrystals for which
$\alpha=1$\cite{Torquato2003}. Other intermediate cases exist (e.g. with
logarithmic scaling), with the term {\em strong hyperuniformity}
reserved for those systems in which $\alpha \ge 1$
\cite{Oguz2017}. Back to our problem, in Figure \ref{s2ncl} we present
the corresponding analysis based on the configurations of the clusters'
center of mass. The number fluctuation is computed by averaging both the
number of clusters, $N_{cl}$ and $N_{cl}^2$ for sample spheres of
radius $R$ randomly placed in the simulation box. Note that particle
number fluctuations are strongly affected by the periodic nature of
our sample, since sampling over r-space is limited in accuracy by
the Fourier space constraint that must be satisfied in the presence of
periodic boundary conditions, namely, ${\bf Q} = (n_x,n_y,n_z)2\pi/L$,
where $n_i$ are integer numbers and $L$ is the side of the simulation
box. For this reason, local number fluctuation can only be accurately
computed for 
values well below half the simulation box size, otherwise effects of the periodic
nature of the sample become visible. These limitations have been
analyzed in detail by Wright in Ref.~\cite{Wright2017}. 
Now, in the left graph of Figure \ref{s2ncl}, one observes that
$\sigma^2_{N_{cl}}(R) 
\propto R^{2}$ except for the high temperature/low density case which
exhibits a $R^3$ dependence. The
curves exhibit strong oscillations, evenly distributed around the
linear regression lines (dotted lines in the figure), and indicate
that the system has strong spatial correlations consistent with a
dense amorphous 
system. On the right graph the scaled ratio
$\sigma^2_N(R)/(4\pi n_{cl}R^3/3)$ is displayed vs the radius of the
sampling sphere, $R$. Here $n_{cl}=<N_{cl}>/L^3$ is  the average
density of clusters. A decreasing trend in the $R$-dependence of this
ratio is a signature of hyperuniformity \cite{Torquato2018a}. Both
graphs in Figure \ref{s2ncl}  indicate that our systems (with
the exception of the low density/high T case) display inter-cluster
structures that can be deemed {\em
  effective hyperuniform} from the point of view of number
(or local density) fluctuations. When the right graph of Figure
\ref{s2ncl} is compared with Figure 2 of
Ref.~\cite{Torquato2018a}, we can clearly appreciate that our systems
are somewhere in between ordered and disordered hyperuniform point
configurations, as could be expected from a glassy-like system. 
The same analysis can be performed on particle number
fluctuations. These results are presented in similar graphs in Figure
S3 of the
Supplementary Information. Here the situation is somewhat
inconclusive. On one hand, the decreasing trend of
$\sigma^2_N(R)/(4\pi \rho R^3/3)$ vs $R$ might be an indication of the
presence of a certain degree of hyperuniformity. On the other, the
strong oscillations on the fluctuation itself (left graph of S3)
prevent to reach a clear conclusion. Most likely, the traces of
hyperuniformity result from the underlying effective hyperuniform
structure of the clusters' centers of mass. 

One might conclude  that from the point of
view of cluster configurations, the systems studied
(with the exception of the high temperature/low density one) 
can be cast into
class I of {\em effectively hyperuniform} materials, since
$\sigma^2_N(R)\sim R^{d-1}$. This corresponds to
$\lim_{Q\rightarrow 0} S(Q) \propto Q^{\alpha}$ \cite{Torquato2018a}
with $\alpha>1$. In  our case the low-Q behavior matches $\alpha\sim
2$, even if the structure factors do not vanish completely in the zero
wavenumber limit. As to the structure displayed by the
particle fluid as a whole, no clear conclusion can be drawn at this
stage.

\section{Conclusions}

In summary, we have presented a study of the structural and dynamic
properties of a simple model of a type-III SALR self-associating fluid,
focusing on the globular cluster phase that upon aggregation displays
a cluster structure with 
effective hyperuniform disorder. Here we have studied both
cluster size distributions, cluster density profiles, diffusion and
frequency spectra as well as pair distributions (inter and
intracluster) and structure factors. From the latter, we have  found that the
cluster phase  for moderate/high cluster densities ($\rho_C\gtrsim 0.8$)
approaches  a cluster glassy state. In these
conditions the cluster structure shows a considerable attenuation of density
fluctuations for a range of wavenumbers below $0.2\sigma^{-1}$,
meeting  the criteria of {\em effective
  hyperuniformity and stealthiness}. When considering the system as a
whole, which includes contributions from free particles, and effects
of cluster size polydispersity, results are less conclusive. Probably
larger samples might throw some light onto this problem, but they are
beyond our current computational capabilities.  One
must bear in mind anyway that a complete study must incorporate the explicit
influence of particle form factors,  which play a fundamental role in
determining the interaction with radiation probes.  Using as starting
point our glassy cluster fluid,
for large colloidal particles ($\sigma\approx 1\sim 10 \mu m$), attenuation would
take place for radiation in the low frequency range of radio
waves. The region of interest  can be tuned resorting to mixtures of
cluster-forming particles 
\cite{Chen2018} and/or controlling the size of the clusters.  Future
work will focus on 
mixtures where intracluster phase 
separation can enable the tuning of anisotropic cluster-cluster
interactions and systems in which self-limitation is controlled by
saturation of associative sites in order to reduce polydispersity.

\section*{Supplementary material}
As supplementary information a graphic illustration of the aggregation
process is  included. In addition, velocity self-correlation
functions are discussed an the net particle number fluctuation
as a function of sampling radius is also introduced to complement the
analysis in the Section V.

 \section*{Acknowledgments}
   The authors acknowledge the support from the Agencia Estatal de
  Investigación and Fondo Europeo de Desarrollo Regional (FEDER) under
  grants No. PID2023-151751NB-I00 and PID2020-115722GB-C22.  We also
  would like to acknowledge the Galicia Supercomputing 
Center (CESGA) for the access to their computer facilities.

\bigskip

%\bibliography{cluster}

 \end{document}